%% file: v4PaperHGW210609.tex
\documentclass[11pt,a4paper,english,twoside]{article}

\usepackage{a4wide}
\usepackage{amssymb, amsmath, amsthm}
\usepackage{graphicx}
\usepackage{subcaption}
\usepackage[all]{xy}
\usepackage{enumerate}
\usepackage[pdftex,hyperref,svgnames]{xcolor}
\usepackage[pdftex,colorlinks=true,
pdfstartview=FitV,
pdfnewwindow=true,
linktoc = page,
linkcolor= Red,
citecolor= blue,
urlcolor= blue,
hyperindex=true,
hyperfigures=false]{hyperref}
\hypersetup{linktocpage}
\usepackage{dsfont}
\usepackage{empheq}
\usepackage{cite}
\usepackage{float}
\usepackage{cancel}
\usepackage{relsize}
\usepackage{soul}
\usepackage{enumitem}
\usepackage{hhline}
\usepackage{multirow}
\usepackage{rotating}
\usepackage{bm}

\newcommand{\beq}{\begin{equation}}
\newcommand{\eeq}{\end{equation}}
\def\bea#1\eea{\begin{align}#1\end{align}}
\def\beal#1\eeal{\begin{subequations}\begin{align}#1\end{align}\end{subequations}}
\newcommand{\nn}{\nonumber}

\renewcommand{\i}{\ensuremath{\textnormal{i}}}

\def\del {\partial}
\def\d {{\rm d}}

\newcommand{\vs}{\bm{\sigma}}

\newcommand{\vm}{\bm{m}}
\newcommand{\vn}{\bm{n}}

\newcommand{\eq}[1]{\begin{equation}#1\end{equation}}

\begin{document}
\numberwithin{equation}{section}

\begin{titlepage}

\begin{center}

\phantom{DRAFT}

\vspace{2.4cm}

{\LARGE \bf{Warp factor and the gravitational wave spectrum}}\\

\vspace{2.4 cm} {\Large David Andriot$^{1}$, Paul Marconnet$^{2}$ and Dimitrios Tsimpis$^{2}$}\\
\vspace{0.9 cm} {\small\slshape $^1$ Institute for Theoretical Physics, TU Wien\\
Wiedner Hauptstrasse 8-10/136, A-1040 Vienna, Austria}\\
\vspace{0.2 cm} {\small\slshape $^2$ Institut de Physique des Deux Infinis de Lyon\\
Universit\'{e} de Lyon, UCBL, UMR 5822, CNRS/IN2P3\\
4 rue Enrico Fermi, 69622 Villeurbanne Cedex, France}\\
\vspace{0.5cm} {\upshape\ttfamily david.andriot@tuwien.ac.at; marconnet@ipnl.in2p3.fr; tsimpis@ipnl.in2p3.fr}\\

\vspace{2.8cm}

{\bf Abstract}
\vspace{0.1cm}
\end{center}

\begin{quotation}
\noindent A distinct signature of compact extra dimensions would be a Kaluza--Klein tower of gravitational waves. Motivated by this prospect, we compute the corresponding spectrum on a warped toroidal background. We evaluate in particular the impact of the warp factor on the spectrum. To that end, we use the complete warp factor $H$ of standard string compactifications, generated by $D$-branes and orientifolds, thus connecting to recent works on stringy de Sitter constructions. The problematic region close to an orientifold where $H<0$ leads to unphysical tachyonic modes in the spectrum. We develop tools that overcome this difficulty and lead to a tachyon-free spectrum. We show, in particular, that the warp factor can lower the first Kaluza--Klein mass by at least $69\%$.
\end{quotation}

\end{titlepage}

\newpage

\tableofcontents

\section{Introduction}

Gravitational waves astronomy has developed in recent years at a remarkable pace. In the latest run (O3) of ground based detectors LIGO and Virgo (lately joined by KAGRA and GEO), one to two candidate events were detected every week \cite{Abbott:2020niy, Abbott:2020gyp}. Further recent or coming observations are promising, including results by NANOGrav (and PTA) \cite{Arzoumanian:2020vkk}, while numerous exciting experiments are planned, among which the awaited eLISA. This impressive new observational window provides non-trivial tests of General Relativity \cite{Abbott:2020jks}, but it also offers the prospect of discovering new physics. Recent reviews on expectations for fundamental physics from gravitational waves can be found e.g.~in \cite{Barausse:2020rsu, Perkins:2020tra, Chia:2020dye, Calcagni:2020ume}. One fascinating example would be evidence for new scalar fields, possibly axions, as described through scalar-tensor models. Such fields could for instance be present in scalar clouds around black holes, thus having various impacts on emitted gravitational waves, or even lead to scalar waves, or effects on black hole shadows \cite{Kuntz:2020yow, Wong:2020qom, Khodadi:2020jij, Gorghetto:2021fsn}. Another interesting example are constraints or predictions that can be made with gravitational waves in string theory frameworks \cite{Addazi:2020obs, Mehta:2020kwu, Abbott:2021ksc, Mehta:2021pwf}. In this work, we are interested in new fundamental physics at high energy, that could leave specific signatures in gravitational waves detectable in future experiments. This is possible if the high energy events are taking place in the early universe. Indeed, the redshift due to cosmological expansion can then lower the frequency of primordial gravitational waves, allowing for an observation by e.g.~eLISA. Typical high energy events in the early universe leading to the emission of such gravitational waves are related to primordial black holes \cite{Hooper:2020evu} or (electroweak) first order phase transitions \cite{Caprini:2015zlo, Caprini:2018mtu, Caprini:2019egz, Eichhorn:2020upj}; others can be found e.g.~in \cite{Kite:2020uix}. Here we are interested in extra dimensions and related Kaluza--Klein towers, typically considered at high energies. Recent studies on their impact on gravitational waves include \cite{Du:2020rlx, Megias:2020vek, Dalmazi:2020xou}. We focus in this paper on the Kaluza--Klein tower of (massive) gravitational waves obtained from certain extra dimensions. We assume that events of high enough energy, possibly in the early universe, have excited the first states of this tower, and that the corresponding emitted primordial gravitational waves could be detected in future experiments. Such a possibility would provide a very distinct signature of extra dimensions. A crucial question is then that of the precise spectrum of these four-dimensional (4d) Kaluza--Klein gravitational waves.

\begin{figure}[h]
\begin{center}
\includegraphics[width=0.6\textwidth]{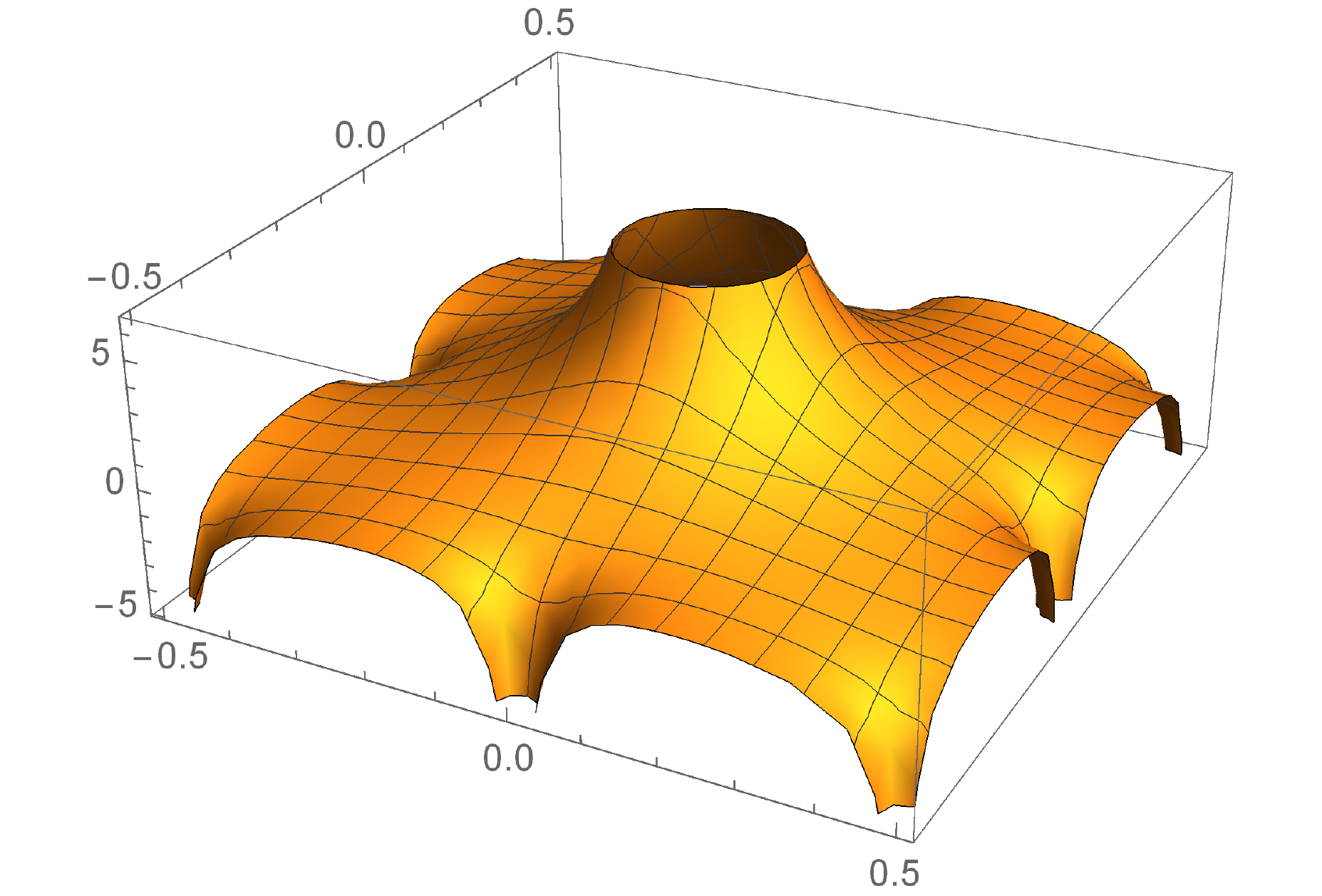}
\caption{Illustration of the warp factor $H$ on a torus $\mathbb{T}^d$, due to the distribution of sources detailed at the beginning of section \ref{sec:towardsspectrum}: $D_p$-branes at the center and $O_p$-planes on the sides. Detailed specifications: $d=3$, $p=6$, $H$ is represented up to a constant and rescaled, i.e.~$g_s^{-1}(L/l_s) (H({\bm \sigma}) - H_0)$, and valued on the vertical axis in terms of the coordinates $\sigma^1, \sigma^2$ on horizontal axis, while $\sigma^3=0$.}\label{fig:H3intro}
\end{center}
\end{figure}

\begin{figure}[h]
\begin{center}
		\resizebox{85mm}{!}{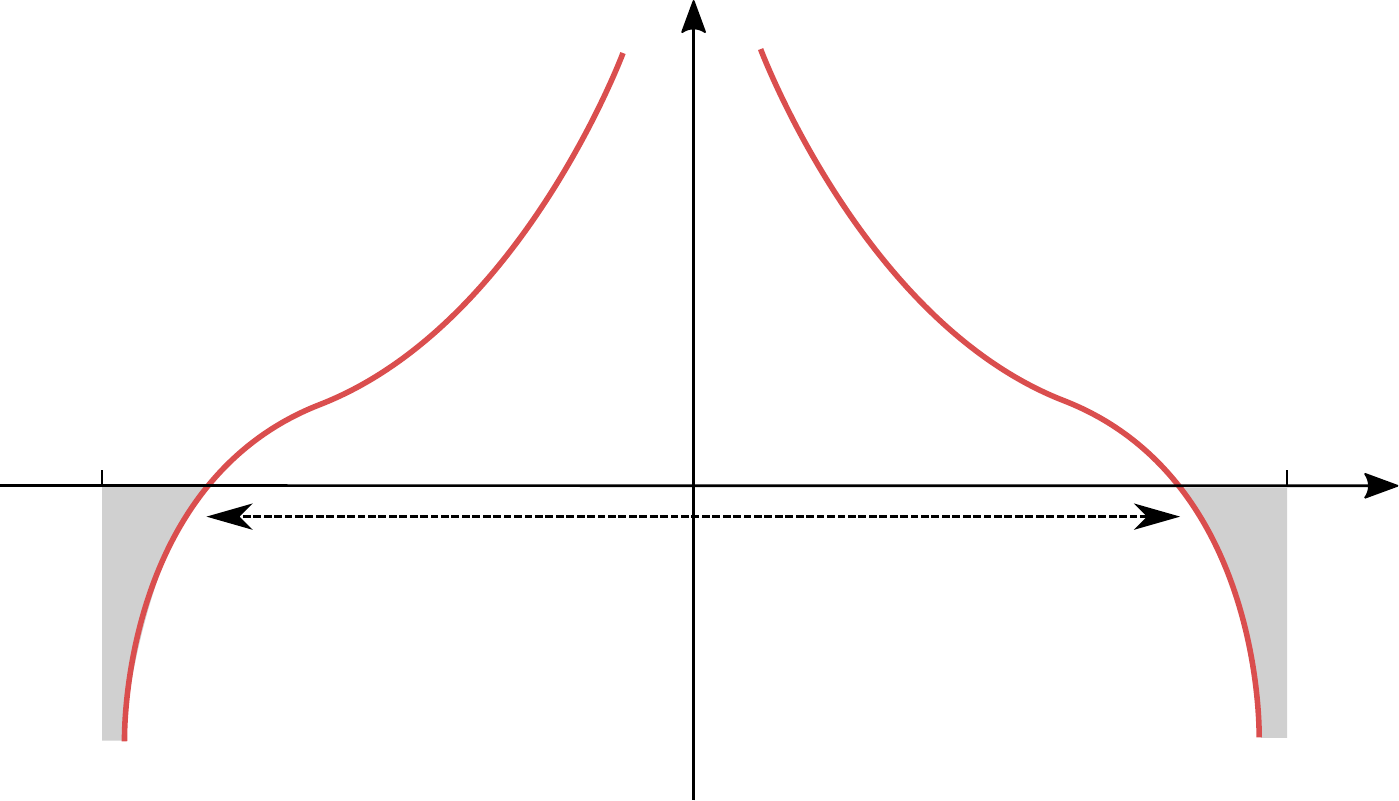}
\caption{Illustration of the warp factor $H$ along one periodic coordinate $\sigma \in (-\tfrac{1}{2},\tfrac{1}{2}]$, generated by $D_p$-branes at $\sigma=0$ and $O_p$ at $\sigma=\frac{1}{2}$. The average of $H$ is adjusted such that $H=0$ at a string length $\sigma=\frac{l_s}{L}$ from the $O_p$. This leaves a ``negative region'', depicted in gray, where $H <0$. To avoid this problematic region, we will conduct the analysis on a restricted domain ${\cal D}$ of size $\lambda=1 - 2\frac{l_s}{L}$. Detailed specifications: $d=2$, $p=7$, the source distribution is given at the beginning of section \ref{sec:towardsspectrum}, $H$ is rescaled as $g_s^{-1} (L/l_s)^{d-2} H$ and plotted along $\sigma^2$, while $\sigma^1=0$.}\label{fig:grapheintro}
\end{center}
\end{figure}

We are interested here in a background with a warp factor. Many bottom-up phenomenological constructions, or top-down models coming from string theory compactifications, include extended objects as branes, that typically host matter and gauge interactions. Such extended objects back-react on the geometry: this is captured in the metric by a function $H$ called the warp factor. It is thus legitimate to consider such a warped gravitational background. Many BSM studies have used backgrounds and warp factor coming from Randall-Sundrum models \cite{Randall:1999ee, Randall:1999vf}. Motivated by a string theory origin, we rather use here the warp factor coming from $p$-brane solutions in supergravity compactifications (see e.g.~\cite{Andriot:2016ufg}). A specificity is that the $D_p$-branes in that case, as well as orientifold $O_p$-planes to be considered, are not only back-reacting on the geometry: they also source a U(1) gauge field (or a generalization thereof). The latter is also described by the warp factor $H$, which is in turn the solution to a Poisson equation, i.e.~a sourced Laplace equation. The warp factor to be considered in such a string or supergravity compactification context is thus not simple, since it involves Green's functions on compact spaces, typically poorly known. We nevertheless tackled this question in \cite{Andriot:2019hay} and provided a complete expression for such a generalized Green's function on a torus $\mathbb{T}^d$, and consequently of the warp factor generated by a distribution of $D_p/O_p$ sources: illustrations are provided in Figures \ref{fig:H3intro} and \ref{fig:grapheintro}. This allowed us to provide a first estimate of the spectrum of Kaluza--Klein gravitational waves, on a background being a warped product of Minkowski and toroidal extra dimensions. This material is reviewed and extended in section \ref{sec:setting}, building on \cite{Bachas:2011xa, Andriot:2017oaz}. In the present work, we will overcome previously unnoticed difficulties and make important technical improvements, that will allow a much more complete and precise determination of this spectrum.\\

The aim of this paper is the determination of the Kaluza--Klein gravitational waves spectrum, but various problems encountered related to the warp factor and its sources could be of broader interest, and find echoes in the recent string compactification literature. To start with, the warp factor has been at the center of many recent discussions, often connected to (anti-) de Sitter compactifications. The warp factor and its determination plays a crucial role in testing some swampland conjectures in specific anti-de Sitter solutions \cite{Junghans:2020acz, Marchesano:2020qvg}. Choosing for it and the dilaton non-standard boundary conditions close to $O_p$ sources has lead to debated new de Sitter solutions \cite{Cordova:2018dbb, Cribiori:2019clo, Cordova:2019cvf, Kim:2020ysx, Bena:2020qpa}; we come back to this discussion in detail in section \ref{sec:away}. As here, though on a different background metric, the impact of the warp factor on the Kaluza--Klein spectrum appeared important in \cite{Blumenhagen:2019qcg}, as it revealed new light states that could play a critical role in the KKLT scenario \cite{Kachru:2003aw}. Kaluza--Klein spectra, in warped compactifications to anti-de Sitter, were also computed recently in e.g.~\cite{Dimmitt:2019qla, Malek:2019eaz, Cesaro:2020piw, Malek:2020yue} (see also \cite{Tsimpis:2012tu, Richard:2014qsa} in relation to scale separation), even though the warp factor is then of different origin and takes a different form. Last but not least, the validity of supergravity approximations has been tested in detail in standard Minkowski compactifications with $O_6$ in \cite{Baines:2020dmu}, building in part on \cite{Andriot:2019hay}.

More generally, this question of supergravity approximations and regime is at the heart of many of the above works, and became dramatically important for the KKLT scenario in the recent works \cite{Gao:2020xqh, Carta:2021lqg}. There, one considers as extra dimensions a warped compact Calabi-Yau manifold, where the warp factor is generated by various $D_p/O_p$. It has been pointed out \cite{Gao:2020xqh} (see also \cite{Carta:2019rhx}) that the negative region around an $O_p$, meaning where $H <0$ as illustrated in Figure \ref{fig:grapheintro}, is likely to be large, i.e.~of order of the size of the Calabi-Yau itself, in order to realise the KKLT scenario. This is problematic because within the used supergravity description, the warp factor is considered positive: see \cite{Andriot:2019hay} or section \ref{sec:backgd} here. A first reason for requiring $H > 0$ is that it enters the metric, and a change of sign would lead to problematic loci. The place where $H=0$ is thus sometimes referred to as the orientifold horizon or the singularity. A physical resolution to this problematic negative region is then hoped to come from string theory: one argues that the classical and perturbative string regime corresponding to supergravity does not hold anymore in that region, and new stringy physics takes over. To enforce this argument in our simple setting, we use the prescription of \cite{Andriot:2019hay} that fixes the horizon at a string length distance from the $O_p$, as depicted in Figure \ref{fig:grapheintro}: this way, we ensure that string degrees of freedom should come in and the supergravity description breaks down. We give more details in section \ref{sec:backgd}, and compare to the compactifications of \cite{Gao:2020xqh} and \cite{Baines:2020dmu}. In \cite{Gao:2020xqh}, other ingredients set the size of this horizon or singularity, forcing it to be large. This implies the need of a stringy description over a large region of the Calabi-Yau. A first sketch of such a possibility has been proposed in \cite{Carta:2021lqg}: it invokes non-perturbative stringy contributions to resolve this problematic region.

This negative region, on top of being unphysical in our supergravity description, became even more problematic to us because it would lead to tachyonic modes in the Kaluza--Klein spectrum. We argue indeed on general grounds in section \ref{sec:neg} and appendix \ref{ap:noncomp} why such a negative region is likely to always generate tachyons. Due to a precision matter discussed in appendix \ref{ap:old}, this phenomenon became more manifest to us when the negative region was large compared to the overall size $L$ of the torus (e.g.~for $L/l_s < 10$). We thus developed tools in section \ref{sec:D} to ignore that region, and solve our spectral problem on a restricted domain ${\cal D}$ where $H > 0$, as illustrated in Figure \ref{fig:grapheintro}, allowing for a spectrum without tachyonic modes. It would be interesting to see whether this problem of tachyons and the tools developed here could apply to the framework of \cite{Gao:2020xqh, Carta:2021lqg}.

The spectrum is determined on the restricted domain ${\cal D}$ where the supergravity description can be trusted. One may wonder whether the remaining region where $H<0$, requiring a stringy description, could lead to important modifications of the spectrum. Capturing this extra contribution from another region actually amounts to fixing boundary conditions for each eigenmode on the restricted domain. This is a standard procedure, used for instance at the horizon of black holes. One may then reformulate the question by asking how much the spectrum is dependent on the boundary conditions imposed in ${\cal D}$. As detailed in section \ref{sec:D}, we consider here periodic boundary conditions on ${\cal D}$: this is actually a generic choice for any (square-integrable, piecewise-continuous) function on an interval. For this reason, we believe that the spectrum determined here is robust. It would be interesting to test the dependence on the boundary conditions more thoroughly. Leaving periodicity on ${\cal D}$ would however require a different resolution method, which goes beyond the scope of this paper.\\

Beyond this treatment of the negative region, several important improvements are made in the numerical method used to solve the eigenmode equation, in comparison to \cite{Andriot:2019hay}. Those are detailed in section \ref{sec:num} and appendix \ref{ap:num}. Various innovations allow to reach a better precision, and many more points (Fourier modes), making use in particular of the hyperoctahedral symmetries of the problem. We get this way a part of the spectrum for $\mathbb{T}^d$ with $d=1, ..., 6$, while we stopped at $d=3$ in \cite{Andriot:2019hay}, and many more eigenmodes for the first dimensions $d$. The spectrum is given in section \ref{sec:spectrum}, and a summary of the results is provided in section \ref{sec:ccl}. Comparison to the older method and (tachyonic) spectrum of \cite{Andriot:2019hay} is made in appendix \ref{ap:old}.

\section{Kaluza--Klein gravitational waves in a warped toroidal background}\label{sec:setting}

Building on \cite{Andriot:2019hay}, we present in section \ref{sec:backgd} the warped background over which Kaluza--Klein gravitational waves are studied, and in section \ref{sec:gwsp} the key equations defining their spectrum. We allow ourselves a digression in section \ref{sec:away}, where we comment on the profile of the warp factor when moving away from a source, connecting to the discussion of \cite{Cribiori:2019clo, Cordova:2019cvf} on boundary conditions.

\subsection{The toroidal $p$-brane background and its warp factor}\label{sec:backgd}

We are interested in Kaluza--Klein gravitational waves propagating in a 4d Minkowski space-time, corresponding to a compactification on a $D$-dimensional toroidal $p$-brane background. In this subsection, we present this background, following \cite{Andriot:2019hay}. The $D$-dimensional background metric is, in Einstein frame,
\beq
\d s_E^2=  H^{-\frac{D-p-3}{D-2}} ( \eta_{\mu \nu}\d x^{\mu}\d x^{\nu} + \delta_{ij}\d x^i\d x^j ) + H^{\frac{p+1}{D-2}} \delta_{mn}\d y^m\d y^n \ . \label{metricEinsteinD}
\eeq
The 4d Minkowski indices are $\mu, \nu = 0, ..., 3$, while the compact toroidal indices are $i,j = 4, ..., p$ and $m,n= p+1, ..., D-1$. The $p$-brane world-volume is along the first $p+1$ dimensions, labeled with $\mu$ and $i$, and it is transverse to the remaining $D-p-1$ dimensions labeled with $m$. The distinction between the parallel and transverse dimensions is made thanks to the warp factor $H$, which has different powers along these different directions. The transverse torus $\mathbb{T}^d$, with $d= D-p-1$, will play a crucial role in the following, because the warp factor only depends on its coordinates $H({\bm y})$. Here and in the following, we denote $d$-dimensional vectors with a boldface, e.g.~${\bm y}$. We will consider a square torus, i.e.~each coordinate verifies the identification $y^m \sim y^m + 2\pi L$ with same radius $L$. Convenient coordinates will then be $\sigma^m = y^m / (2\pi L) \in\ (-\tfrac{1}{2},\tfrac{1}{2}]$.

The $p$-brane background is a solution to an Einstein-Maxwell-dilaton type of theory. The warp factor captures the back-reaction of the $p$-brane, but it also gives the dilaton, and the flux or field-strength sourced electrically (or magnetically, according to conventions) by the $p$-brane. The latter gets translated into a Poisson equation on $H$ over the (unwarped) compact transverse torus $\mathbb{T}^d$, whose solution is given by a generalized Green's function $G$. In the following we will consider not only one but a distribution of such sources, each of them having a charge $Q_i$, and placed at a position ${\bm y}_i$ in $\mathbb{T}^d$. More precisely, in reference to the corresponding stringy objects, the sources will be named $D_p$-brane or orientifold $O_p$-planes, with $D$-dimensional charge
\beq
Q_{D_p} = - (2\pi l_s)^{d-2} g_s \ ,\ Q_{O_p}= - 2^{4-d}\, Q_{D_p} \ , \label{charges}
\eeq
where $l_s$ is the string length (or the fundamental length in a broader setting), and $g_s$ a constant related to the string coupling. In such a distribution of sources, the resulting warp factor was shown in \cite{Andriot:2019hay} to be given by
\beq
\label{H}
H = \sum_i Q_i\, G({\bm y} - {\bm y}_i)  + H_0 \ ,
\eeq
with a constant $H_0$. Both the generalized Green's function $G$ and the constant $H_0$ are non-trivially determined, as we will now recall. We consider in the following a chargeless source configuration, i.e.~with $\sum_i Q_i = 0$: in the absence of extra fluxes, this vanishing is required by compactness \cite{Andriot:2019hay}.\\

The generalized Green's function $G$ on the torus $\mathbb{T}^d$ was discussed and studied in \cite{Andriot:2019hay}, using Courant-Hilbert \cite{Courant:1989aa}, or the comparatively recent \cite{Shandera:2003gx} based as well on century-old mathematical references. Both $H$ and $G$ have to be periodic on $\mathbb{T}^d$, so a first naive expression as a Fourier series is
\beq
G({\bm \sigma}) = -\frac{1}{(2\pi L)^{d-2}} \sum_{{\bm n}\in \mathbb{Z}^{d\, *}} \frac{e^{2\pi \i {\bm n}\cdot {\bm \sigma}}}{4\pi^2 {\bm n}^2} \ , \label{GFourier}
\eeq
with $\mathbb{Z}^{d\, *}$ being $\mathbb{Z}^{d}$ without ${\bm 0}$. This sum is however not absolutely convergent for $d \geq 2$. An appropriate regularization is provided thanks to the following expression
\beq
G({\bm \sigma}) =(2\pi L)^{2-d} \int_0^\infty \d t \Big( 1-\prod_{m=1}^d\theta_3(\sigma^m|4\pi \i t) \Big) \ , \label{Gtheta}
\eeq
in terms of the theta function $\theta_3= \theta_{00}$.\footnote{The convention here is
\beq
\theta_3(\sigma|\tau)=\sum_{n \in \mathbb{Z}} {\rm e}^{2\pi \i(n\sigma +\frac{n^2}{2}\tau)} = 1 + 2 \sum_{n=1}^\infty q^{n^2} \cos(2\pi n \sigma ) \ ,\ {\rm with}\ q= {\rm e}^{\i \pi \tau}
~.\nn
\eeq} This expression holds up to a constant, which will not matter in $H$ thanks to $\sum_i Q_i = 0$. Studying this expression in \cite{Andriot:2019hay}, we could recover analytically the expected behaviour close to the source, namely
\bea
d\geq 3 &:\quad  (2\pi L)^{d-2}\ G({\bm \sigma}) \sim_{{\bm \sigma}^2 \rightarrow 0}\  - \frac{1}{4 \pi^{\frac{d}{2}}}\ \Gamma\left(\tfrac{d-2}{2} \right)\, \frac{1}{|{\bm \sigma}|^{d-2}} \ , \nn\\
d=2 &:\quad G({\bm \sigma}) \sim_{{\bm \sigma}^2 \rightarrow 0}\ \frac{1}{2 \pi} \ln |{\bm \sigma}| \ , \label{behiaviourG}\\
d=1 &:\quad (2\pi L)^{-1}\ G({\bm \sigma}) \sim_{{\bm \sigma}^2 \rightarrow 0}\ -\frac{1}{12} + \frac{|{\bm \sigma}|}{2} \ . \nn
\eea

The constant $H_0$ is a crucial piece of information. One first shows that it is the average of $H$: for $ \sum_i Q_i = 0$, using the periodicity in $\sigma^m$, one gets
\beq
\int_{-\frac{1}{2}}^{\frac{1}{2}} \d^d {\bm \sigma}\ H = H_0 \ .\label{intH}
\eeq
One also verifies thanks to \eqref{behiaviourG} that the $d$-dimensional integral of $G$ is finite. In the case where $ \sum_i Q_i \neq 0$, one could add to $G$ a constant opposite to its $d$-dimensional integral, allowing to recover \eqref{intH}. We show additionally in \cite{Andriot:2019hay} that this average of $H$ \eqref{intH} appears in several important places, including the 4d Planck mass, or a condition necessary to have a standard massless 4d gravitational wave. It is then required to have $H_0 \neq 0$. Its value actually plays an important role, as we now explain.

To avoid singularities and signature changes in the background metric \eqref{metricEinsteinD}, we require $H>0$. As indicated with Figure \ref{fig:grapheintro}, the source distribution considered however typically generates a part of the space where $H<0$. This happens close to the $O_p$. Such sources remain necessary on a compact space because they provide charges opposite to those of the $D_p$. A strategy then consists in controlling the size of this ``negative region'' where $H<0$, by allowing it to be at most of string length size away from the $O_p$. The reason is that below a distance $2 \pi l_s$, new stringy physics is expected while supergravity description breaks down, so our analysis can in any case only be trusted up to this point. Fixing this distance is made possible thanks to the constant $H_0$, which ``shifts $H$ vertically'' in Figure \ref{fig:grapheintro}. The corresponding prescription of \cite{Andriot:2019hay} was
\beq
H_0 =  - \min_j \, \left\{\sum_i Q_i\, G({\bm \sigma} - {\bm \sigma}_i)|_{|{\bm \sigma} - {\bm \sigma}_j|=\frac{ l_s}{L}} \right\} \ , \label{H0prescr}
\eeq
such that $H=0$ at a distance $|{\bm y} - {\bm y^j}| = 2 \pi l_s$, i.e.~$|{\bm \sigma} - {\bm \sigma}_j|=\frac{ l_s}{L}$, of any $O_p$ labeled $j$ or closer to it. Note that in each circle of $\mathbb{T}^d$, a distance of $2\pi l_s$ at least should be allowed on each side of an $O_p$, to be able to have a further ``positive'' region described by supergravity. So circle perimeters should be larger, i.e.~$2 \pi L > 2 \times 2\pi l_s$ or $L/l_s > 2$. Finally, in the approximation $L \gg l_s$, concrete values for $H_0$ were computed from \eqref{H0prescr} in \cite{Andriot:2019hay} for $d\geq 2$, using \eqref{behiaviourG}, namely
\beq
d \geq 3:\ H_0 = g_s\, 2^{2-d} \pi^{-\frac{d}{2}}\ \Gamma\left(\tfrac{d-2}{2} \right) \ , \quad d=2: \ H_0 = g_s\, \frac{2}{\pi} \ln \left(\frac{L}{l_s}\right) \ .\label{constantH0}
\eeq

Few comments are in order regarding the prescription \eqref{H0prescr}. Relative to $\eta_{\mu\nu}$, the metric of the transverse torus is proportional to $ H 4 \pi^2 L^2$.\footnote{From this observation, one could argue that a proper evaluation of a ``string length distance'' from an $O_p$ for the prescription \eqref{H0prescr} should include the warp factor, or at least $H_0$. This makes the determination of $H_0$ more complicated, and for simplicity we stick to the prescription as stated. The method developed in this paper to determine the gravitational waves spectrum can in any case be adapted to a different value of $H_0$.} One may then view $\sqrt{H}$, or at least $\sqrt{H_0}$, as part of the physical radius, contributing to the volume of these compact dimensions; we come back to this point in section \ref{sec:gwsp}. The resulting ambiguity between $L$ and $\sqrt{H_0}$ was lifted in \cite{Gao:2020xqh} by completely fixing $L$, and letting $H_0$ capture volume fluctuations. On the contrary, $H_0$ was set to 1 in \cite{Baines:2020dmu} to match the smearing conventions, leaving the freedom to $L$, and verifying in that case the validity of supergravity approximations. The prescription \eqref{H0prescr} is yet another option. Whatever choice is made, the volume or corresponding radius, as a 4d scalar field, could be stabilized at a given value by further physical ingredients generating an appropriate potential. These are precisely physical requirements (on Euclidian instantons), necessary to realise the KKLT scenario, that fix in \cite{Gao:2020xqh} $H_0$ to a low value. Our compactification setting is much simpler, and although it could be interesting to study the effective 4d action using e.g.~\cite{Giddings:2005ff, Shiu:2008ry, Douglas:2008jx, Frey:2008xw}, we will leave here the volume and radius unfixed. Therefore, we do not consider any further constraint than \eqref{H0prescr} on $H_0$ and $L$, and will analyse the outcomes for various values of $L$.

\subsection{Apart{\'e}: moving away from a source}\label{sec:away}

\begin{figure}[H]
\begin{center}
\begin{subfigure}[H]{0.55\textwidth}
\includegraphics[width=\textwidth]{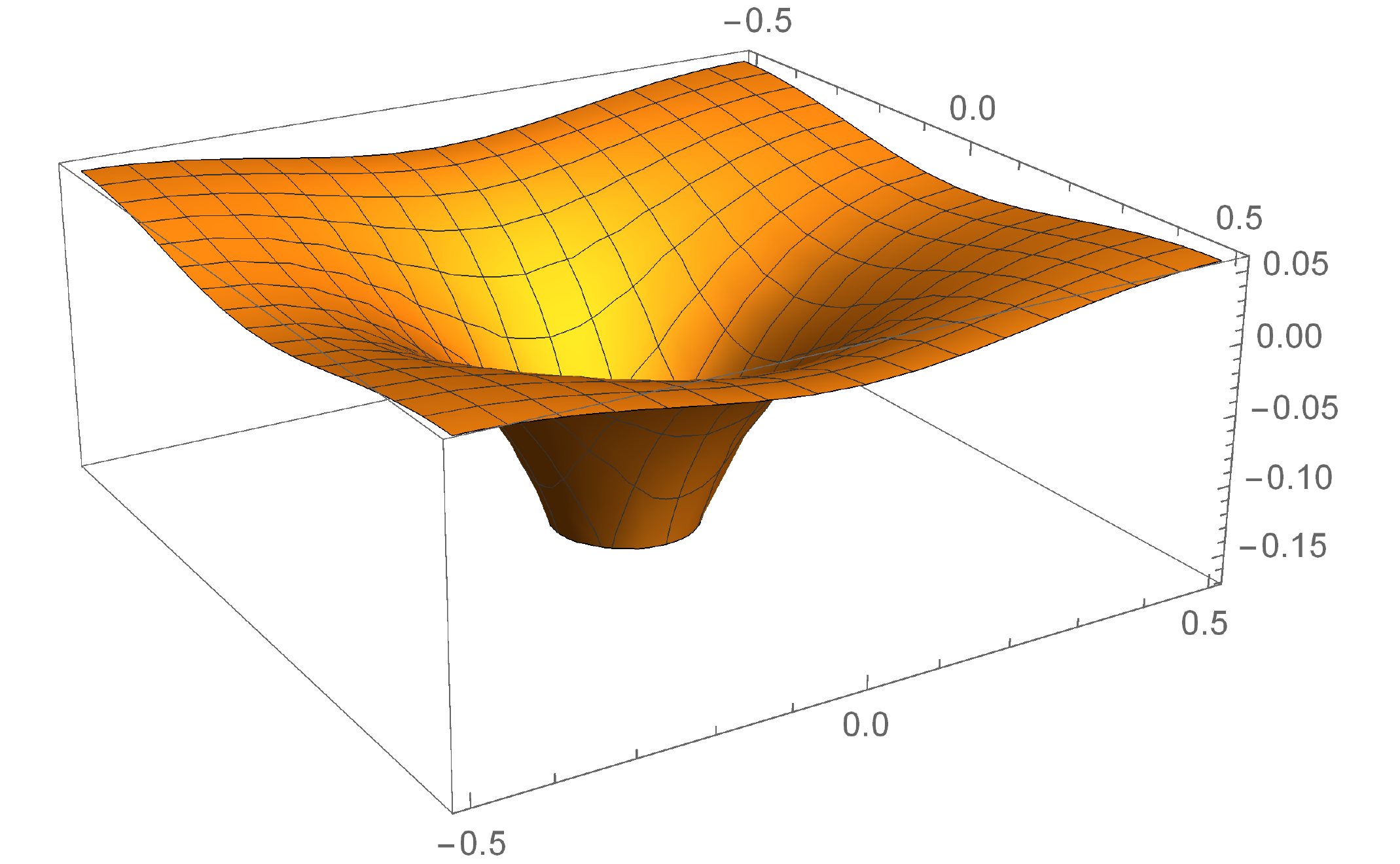}\caption{$G({\bm \sigma})$}\label{fig:G2}
\end{subfigure}
\quad
\begin{subfigure}[H]{0.35\textwidth}
\includegraphics[width=\textwidth]{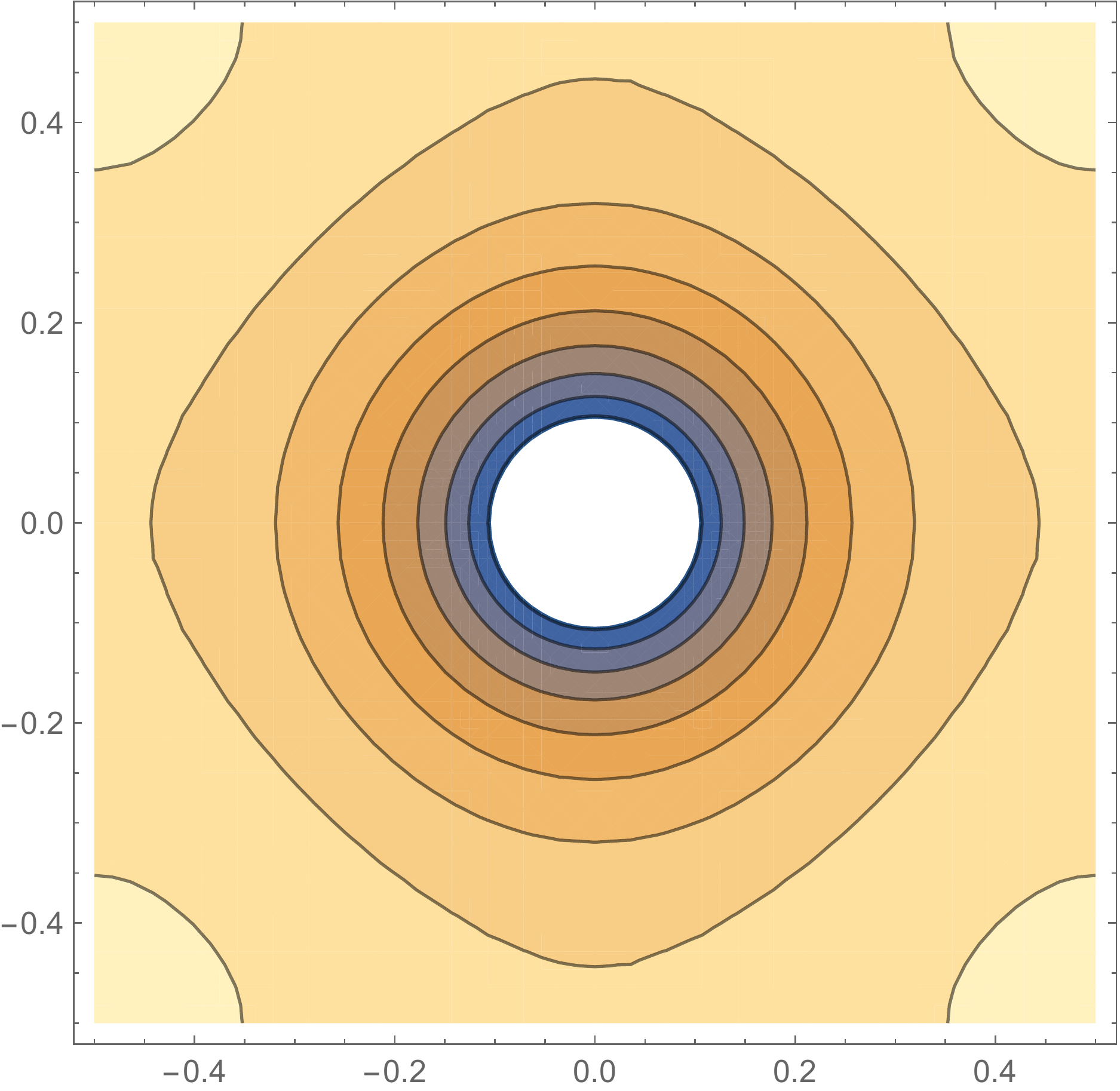}\caption{Figure \ref{fig:G2} seen from above}\label{fig:G2cut}
\end{subfigure}
\begin{subfigure}[H]{0.06\textwidth}
\includegraphics[width=\textwidth]{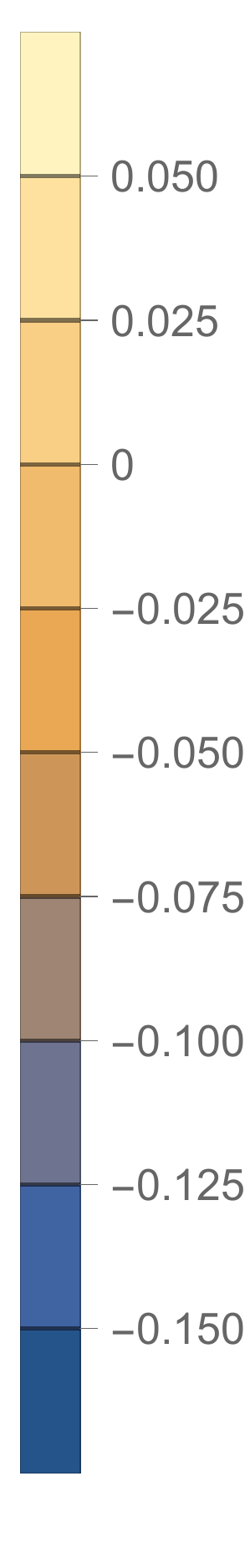}\label{fig:G2cutscale}
\end{subfigure}
\caption{Green's function $G({\bm \sigma})$ given in \eqref{Gtheta} for $d=2$, evaluated on the vertical axis in Figure \ref{fig:G2} in terms of the horizontal $\sigma^{1}, \sigma^2$ coordinates. The same is displayed from above in Figure \ref{fig:G2cut}, with horizontal cuts at fixed values of $G({\bm \sigma})$. Close to the source (at ${\bm \sigma}={\bm 0}$) we verify the spherical symmetry, but it is broken further away, where the circle turns to an approximate square.}\label{fig:symG2}
\end{center}
\end{figure}

\begin{figure}[H]
\begin{center}
\begin{subfigure}[H]{0.4\textwidth}
\includegraphics[width=\textwidth]{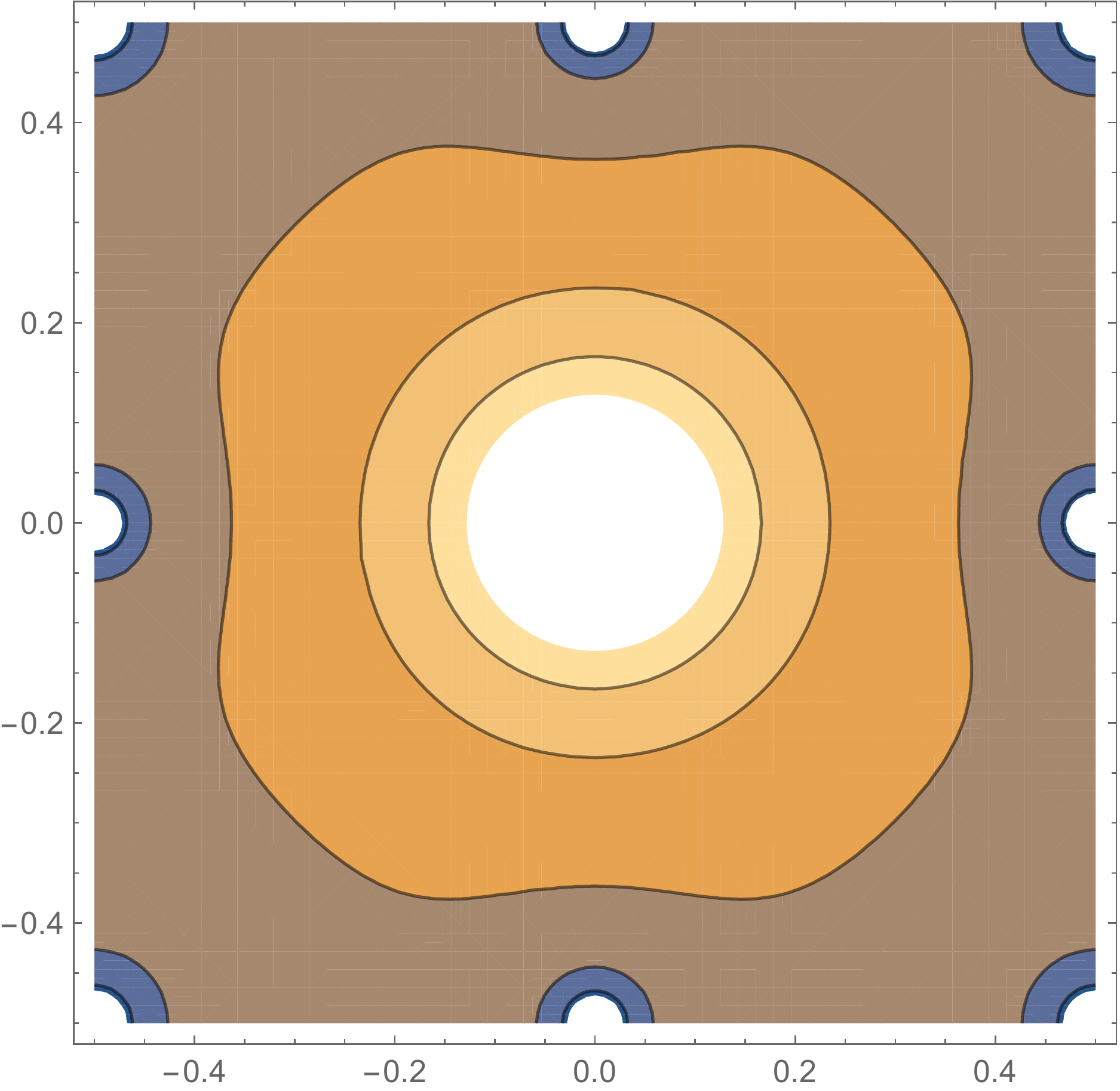}\caption{Figure \ref{fig:H3intro} seen from above}\label{fig:H3cut}
\end{subfigure}
\begin{subfigure}[H]{0.045\textwidth}
\includegraphics[width=\textwidth]{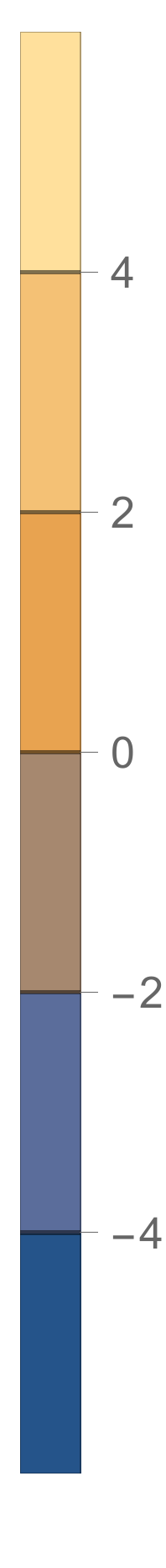}\label{fig:H3scale}
\end{subfigure}
\qquad \qquad
\begin{subfigure}[H]{0.4\textwidth}
\includegraphics[width=\textwidth]{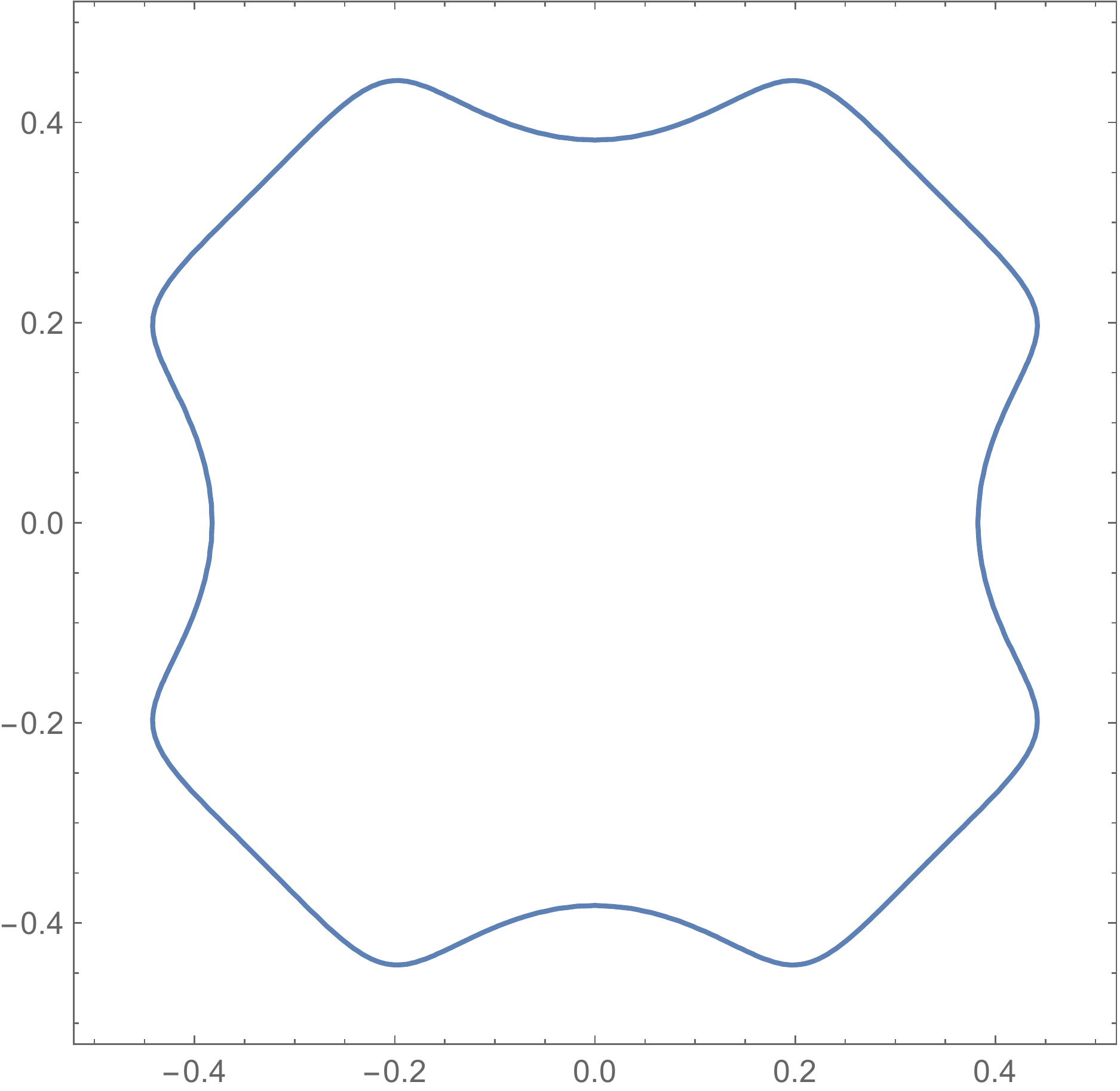}\caption{Horizontal cut at $-0.3$}\label{fig:H3cutnb}
\end{subfigure}
\caption{Warp factor $H$ for $d=3$ displayed in Figure \ref{fig:H3intro}, viewed from above. Figure \ref{fig:H3cut} and \ref{fig:H3cutnb} show horizontal cuts at fixed values of $H$ (several values for \ref{fig:H3cut}, one value for \ref{fig:H3cutnb}). The breaking of spherical symmetry around sources is made explicit by contours evolving from circles to angular shapes. Similar illustrations can be obtained for $d=2$.}\label{fig:Hcut}
\end{center}
\end{figure}

We make here side comments on the profile of the Green's function and the warp factor when moving away from the sources, the related symmetries and boundary conditions. Close to a source, the generalized Green's function on $\mathbb{T}^d$ exhibits a spherical symmetry as can be seen in \eqref{behiaviourG}. For $d \geq 2$, this symmetry is however broken further away from the source: there is indeed no $\text{SO}(d)$ symmetry among coordinates $\sigma^m$ in the complete expression \eqref{Gtheta}. Showing this analytically, following appendix A of \cite{Andriot:2019hay}, and getting a coordinate-dependent correction to the spherical behaviour \eqref{behiaviourG}, turns out to be difficult. The breaking of this symmetry can nevertheless be verified numerically, as displayed in Figure \ref{fig:symG2}.
In the warp factor $H({\bm \sigma})$, the spherical symmetry around sources gets broken for an additional reason: the presence of other sources. This is made manifest in Figure \ref{fig:Hcut}. This point highlights the need to use the complete expressions of $G$ and $H$, instead of only \eqref{behiaviourG}. This is important when evaluating $H$, to compute the constant $H_0$ as proposed with the prescription \eqref{H0prescr}. The computation of $H_0$ in \eqref{constantH0} rather made use of the spherical symmetry, valid there only thanks to $l_s/L \ll 1$.

A related question has been discussed in \cite{Cribiori:2019clo, Cordova:2019cvf}, following new 10d supergravity de Sitter solutions obtained in \cite{Cordova:2018dbb} with a non-standard ansatz. There, the metric exhibits several different warp factors, to describe $O_{8_+}/ O_{8_-}$ orientifolds transverse to a circle of coordinate $y$ (i.e.~$d=1$ here). We simplify here the discussion by restricting the setting considered there to the present ansatz. We take the $O_{8_+}$ to correspond to a stack of $D_8$ (with one $O_8$) located at $y=0$, and the $O_{8_-}$ to be an $O_8$ located $y_0$: this matches the source configuration considered here, detailed at the beginning of section \ref{sec:towardsspectrum}. The focus will be on the $O_8$ at $y_0$. The ansatz of \cite{Cordova:2019cvf} boils down to ours if the three functions considered there, $W,\, \lambda,\, \phi$, verify the following relations to our warp factor: $e^{-4W}=H ,\, e^{\phi} = g_s H^{-\frac{5}{4}} ,\, \lambda= 2 W$. Those functions are discussed there by related functions $f_{i=1,2,3}$. Their first derivative with respect to $y$ verifies, in our setting, $f_i'= W'$ $\forall i$.

The discussion of \cite{Cordova:2019cvf} is that of the boundary conditions to impose on the various functions close to the $O_8$ at $y_0$. Two possible boundary conditions are put forward: the permissive ones and the restrictive ones. We translate them here in our setting
\bea
&{\rm Permissive:}\quad H'|_{y\rightarrow y_0^+} = \frac{Q_{O_8}}{2} \ ,\\
&{\rm Restrictive:}\quad \frac{H'}{H}\Big|_{y\rightarrow y_0^+} = \frac{Q_{O_8}}{2} \frac{1}{H}\Big|_{y\rightarrow y_0^+} \ ,\label{restrH}
\eea
where we take $H \geq 0$. As we will see, it turns out that the restrictive boundary conditions require one condition more than the permissive ones, so the question is: which boundary conditions should be imposed? To see the difference, one considers the following expansion for $H$ (and more generally for exponentials of the functions $f_i$ in \cite{Cordova:2019cvf})
\beq
H(y)= a_1\, |y-y_0| + a_2\, |y-y_0|^2 + {\cal O}(|y-y_0|^3) \ . \label{Hexp}
\eeq
This means that $H(y=y_0)= 0$, which is always possible by adjusting the constant $H_0$, and this property actually holds in the $d=1$ warp factor considered in \cite{Andriot:2019hay} (see Figure 3); in \cite{Cordova:2019cvf} this is a property of the solution considered. In addition, the expansion \eqref{Hexp} is precisely about the above discussion: the profile of the warp factor when going away from the source. Indeed, the leading behaviour $|y-y_0|$ is guaranteed by the Green's function close to the source \eqref{behiaviourG}, and the subleading terms in \eqref{Hexp} are possible corrections to it. The permissive boundary conditions give one condition only, on the leading term, while the subleading terms all give vanishing contributions at $y_0$. On the contrary, the restrictive boundary conditions require to look at two terms in the Taylor expansion, the pole and the constant one, the remaining ones vanishing at $y_0$. This expansion for the restrictive boundary conditions is captured by the following in \cite{Cordova:2019cvf} (for each function $f_i$)
\beq
{\rm Restrictive:}\quad \frac{d_i^L}{|y-y_0|} + e_i^L  = \frac{d_i^R}{|y-y_0|} + e_i^R \ \Longleftrightarrow \ d_i^L = d_i^R \ , \ e_i^L = e_i^R \ ,
\eeq
where the superscripts $L,R$ refer to each side of the equation, here given by \eqref{restrH}. With the other boundary conditions, only one coefficient is fixed
\beq
{\rm Permissive:}\quad d_i^L = d_i^R \ .
\eeq
The choice of boundary conditions has far reaching consequences in the context of \cite{Cordova:2018dbb, Cribiori:2019clo, Cordova:2019cvf}. Indeed, \cite{Cribiori:2019clo} considered restrictive boundary conditions and generalizations thereof, and deduced a no-go theorem against de Sitter solutions, thus potentially in contradiction with the solutions of \cite{Cordova:2018dbb}. The latter however verified only specific permissive boundary conditions, while violating the restrictive boundary conditions, given that in their solutions, $e_i^L \neq e_i^R$.

In our setting, this difference is even more dramatic. Indeed, considering only the warp factor $H$, expanded as in \eqref{Hexp}, we compute both boundary conditions and get (up to higher powers of $|y-y_0|$, which vanish at $y_0$)
\bea
&{\rm Permissive:}\quad a_1 = \frac{Q_{O_8}}{2}  \ ,\\
&{\rm Restrictive:}\quad \frac{1}{|y-y_0|} + \frac{a_2}{a_1} = \frac{Q_{O_8}}{2 a_1} \left( \frac{1}{|y-y_0|} - \frac{a_2}{a_1} \right) \ \Longleftrightarrow \ a_1 = \frac{Q_{O_8}}{2} \ , \ a_2=0 \ .
\eea
In other words, the restrictive boundary conditions do not allow for a correction in $|y-y_0|^2$ in $H$ when moving away from the source. This is in line with the fact that solutions of \cite{Cordova:2018dbb} do not satisfy those boundary conditions. This was made more explicit in \cite{Kim:2020ysx} which indicates precisely $|y-y_0|^2$ corrections in the various functions.

In the present context, given the complete expressions we have for the Green's function and warp factor, we could attempt, as mentioned above for $d\geq 2$, to determine the corrections away from the source for $d=1$. Interestingly, for $d=1$, the Fourier series \eqref{GFourier} provides an exact expression of the Green's function. This allows us to find an alternative expression valid on one interval $\sigma \in (-\frac{1}{2}, \frac{1}{2}]$,\footnote{We thank D.~Junghans for pointing to us the possibility of such an expression.} by identifying the Fourier coefficients: we give it here in the last equality
\beq
(2\pi L)^{-1}\ G(\sigma) = \int_0^\infty \d t \Big( 1- \theta_3(\sigma|4\pi \i t) \Big) = - \sum_{n\in \mathbb{Z}^{*}} \frac{e^{2\pi \i n \sigma}}{4\pi^2 n^2} = -\frac{1}{2} \left(\sigma^2 - |\sigma| + \frac{1}{6} \right) \ , \label{G1}
\eeq
where again, the first two expressions are periodic while the last one is only valid on one interval, and should be mirrored on other intervals. It is interesting that the last expression for $G(\sigma)$ captures {\it all} corrections away from the source. This result is in agreement with the warp factor we already identified around Figure 3 in \cite{Andriot:2019hay}: we obtained there the following complete warp factor
\beq
H(y) = \frac{Q_{O_8}}{2} (y_0 -|y|) \ ,\ y\in (-y_0, y_0] \ \leftrightarrow\ H(y) = \frac{Q_{O_8}}{2} |y-y_0| \ ,\ y\in (0, 2y_0] \ ,
\eeq
with $y_0=\pi L$. In other words, the warp factor found for $d=1$ with our source configuration has {\it only} the behaviour close to the sources, without correction, contrary to the other dimensions $d$ as mentioned above. In particular, the quadratic terms of \eqref{G1} drop out by charge cancelation, the requirement of $a_2=0$ is verified, and so are the restrictive boundary conditions.\footnote{The fact our setting verifies the restrictive boundary conditions might be expected from \cite{Cribiori:2019clo}. There, it is argued that similar boundary conditions should be obtained whenever one works with the standard DBI + WZ action for sources. The latter holds for us, given our background is a standard type II Minkowski solution.} This is also the case of standard warp factors for $D_8/O_8$ in a (non-compact) Minkowski space-time, as mentioned in \cite{Cordova:2019cvf}. In that respect, the de Sitter solutions of \cite{Cordova:2018dbb} are certainly different.

While this choice of boundary conditions is connected to other interesting questions in \cite{Cordova:2019cvf}, one point is particularly emphasized: the failure of the supergravity description close to the source. As mentioned already, considering short distances could involve string scale physics, and thus a break-down of the supergravity description. This is even more manifest here as $e^{\phi}$ diverges close to the $O_8$ since the warp factor vanishes, so one argues that the string coupling cannot be considered weak anymore. The supergravity equations that define the solutions, the warp factor and its expansion, may then be disputed. This brings us back to the idea of considering a string-length distance away from the source, as discussed around the prescription \eqref{H0prescr}. We will come back to this idea in section \ref{sec:D}.

\subsection{Gravitational waves and their spectrum}\label{sec:gwsp}

We are interested in 4d gravitational waves propagating on the background of section \ref{sec:backgd}, so we consider the fluctuations
\beq
\eta_{\mu\nu} \rightarrow \eta_{\mu\nu} + h_{\mu\nu} \ ,
\eeq
where $h_{\mu\nu}$ depends a priori on all $D$ coordinates. One then decomposes it as a Kaluza--Klein tower of 4d gravitational waves, each mode being generically labeled by $N$
\beq
h_{\mu\nu} = \sum_N h_{\mu\nu}^N(x^{\mu})\ \psi_N (y^m) \ .
\eeq
We could add a dependence of $\psi_N$ on the other extra coordinates $x^i$ but as shown in \cite{Andriot:2019hay}, those will not play any role, especially for toroidal directions. The modes $ h_{\mu\nu}^N$ are taken transverse and traceless in 4d; this can be viewed as a consistent truncation \cite{Andriot:2019hay, Andriot:2017oaz}. Provided the $\psi_N$ are orthonormal eigenfunctions of a certain modified Laplacian operator, to be specified, with eigenvalues $M_N^2$, then each mode satisfies the Pauli-Fierz equation of a massive spin-2 field with mass $M_N$ in Minkowski
\beq
\left(\eta^{\kappa\lambda}\partial_{\kappa}\partial_{\lambda} - M_N^2 \right)h^{N}_{\mu\nu} = 0 \ ,
\eeq
at linear order. This was shown in \cite{Bachas:2011xa} to hold for any energy-momentum tensor, i.e.~any matter content of the theory, thanks to having a maximally symmetric 4d background space-time. The generality of these equations, describing propagating 4d Kaluza--Klein gravitational waves on a warped Minkowski background, is thus interesting. A generalization of this setting was considered in \cite{Andriot:2017oaz} allowing fluctuations of the full $D$-dimensional metric, leading to additional 4d vector and scalar contributions with interesting effects. We refer to \cite{Andriot:2019hay} for more details.

As shown in \cite{Andriot:2019hay}, on the background metric \eqref{metricEinsteinD} with the transverse torus $\mathbb{T}^d$, the modified Laplacian operator and corresponding eigenmode equation boil down to
\beq
-\delta^{mn} \frac{\partial}{\partial\sigma^m}\frac{\partial}{\partial\sigma^n} \psi_N=(2\pi L\, M_N)^2\, H\,  \psi_N \ . \label{psieq}
\eeq
In absence of any source, the warp factor is given by its constant part, $H=H_0$. As explained at the end of section \ref{sec:backgd}, we then simply face a torus of radius $\sqrt{H_0} L $. The Kaluza--Klein spectrum in that case is the standard one: the masses are
\beq
\left(M_N^{{\rm (st)}} \right)^2 = \frac{N^2}{H_0 L^2} \ . \label{st}
\eeq
This is precisely what we recover from \eqref{psieq}, writing $\psi_N$ as a Fourier series on $\mathbb{T}^d$, with $N=|{\bm n}|,\ {\bm n} \in \mathbb{Z}^d$. The variation of $H$ beyond this average, due to the presence of $D_p$ and $O_p$ sources, makes the equation \eqref{psieq} much more complicated to solve. The main purpose of this work is to determine how much the spectrum deviates from the standard one \eqref{st} in presence of a (non-trivial) warp factor $H$. To that end, few techniques were introduced in \cite{Andriot:2019hay}, and they will be greatly improved in the following. Finally, let us recall that
\beq
d \geq 2\ ,\ \ \frac{L}{l_s} \gg 1 \quad \Rightarrow \quad M_N \approx M_N^{{\rm (st)}} \ , \label{matchst}
\eeq
while we stated below \eqref{H0prescr} the minimal value: $L/l_s > 2$. Deviations from the standard spectrum are thus expected close to this last bound.

\section{Issues and method to determine the spectrum}\label{sec:towardsspectrum}

We present here the method, both analytical and numerical, used to determine the Kaluza--Klein gravitational waves spectrum, defined in section \ref{sec:setting}. The spectrum is governed by the eigenmode equation \eqref{psieq}, while the reference for this spectrum is the standard one obtained in the absence of sources, i.e.~with a trivial warp factor, given in \eqref{st}. To determine the spectrum, we first need to overcome difficulties due to the negative region where $H <0$ (see Figure \ref{fig:grapheintro}), responsible for tachyons. This is discussed in sections \ref{sec:neg} and \ref{sec:D}. The numerical method is then presented in section \ref{sec:num} and appendix \ref{ap:num}.\\

Prior to determining the spectrum, we first have to fully specify the background, by indicating where we place our sources, i.e.~give the vectors ${\bm \sigma}_i$ (or ${\bm y}_i$). From now on, we consider the $D_p$-branes to be all at the origin in coordinates ${\bm \sigma}$ (or ${\bm y}$). The orientifolds $O_p$ are at the 2 fixed points of each circle of $\mathbb{T}^d$: those are at $\sigma^m=0$ or $\frac{1}{2}$. There are thus $2^d$ distributed $O_p$. As specified in section \ref{sec:backgd}, we take an overall vanishing charge $\sum_i Q_i = 0$. Since the charge ratio \eqref{charges} is given by a factor $2^{4-d}$, this always gives 16 $D_p$ at the origin. Note that one $O_p$ also sits at the origin, making the total charge there slightly less negative. The positive charges, that give a negative $H$, are then at all the other positions of the $O_p$, as illustrated in Figure \ref{fig:grapheintro}. This charge distribution exhibits certain discrete symmetries, which will be very helpful to the numerical resolution, as described at the end of section \ref{sec:num}. We now enter the details of the determination of the spectrum.

\subsection{Negative region and tachyons}\label{sec:neg}

To determine the spectrum, we first improved the (numerical) method presented in \cite{Andriot:2019hay}, as detailed in section \ref{sec:num}. The resulting spectrum for $d=1,2,3$ is given in appendix \ref{ap:old}. Doing so, we however noticed the systematic presence of tachyons at low $L/l_s$, i.e.~eigenmodes with $M_N^2 < 0$. We understood that those are due to the negative region, meaning the region discussed in the Introduction and section \ref{sec:backgd} where $H<0$. This can be seen in several ways. First, it was noticed in \cite{Andriot:2019hay} (see e.g.~Figure 2 or section 4.2) that low $L/l_s$ make the variation of $H$ stronger and the negative region larger, so the impact of this region could then be more important. Secondly, the relation between tachyons and the negative region is most easily seen using the eigenmode equation \eqref{psieq}, as follows
\beq
0 \leq \int_{-\frac{1}{2}}^{\frac{1}{2}} \d^d {\bm \sigma}\ |\del \psi_N|^2 = - \int_{-\frac{1}{2}}^{\frac{1}{2}} \d^d {\bm \sigma}\ \psi_N^* \del^2 \psi_N   =  (2\pi L\, M_N)^2 \int_{-\frac{1}{2}}^{\frac{1}{2}} \d^d {\bm \sigma} \ H \,  |\psi_N|^2 \ . \label{intHtachyon}
\eeq
For a constant $H$, we deduce that $H\, M_N^2 \geq 0$. A negative constant $H$ then leads to a tachyon. More generally, we infer that if the negative region is sufficiently large, as at low $L/l_s$, it may dominate the above integral, at least for some mode, and a tachyon can appear. Similarly, probing the negative region may require a small enough wavelength of the eigenmode, which may then lead to a negative integral \eqref{intHtachyon}, making the mode tachyonic. Small wavelengths correspond to Kaluza--Klein modes high in the tower. Those could then be truncated by our numerical approach, that only considers a finite number of modes. At low $L/l_s$, the negative region is larger, and such modes are more easily reached. This may explain why we only noticed tachyons at low $L/l_s$, while they may always exist as long as $H<0$ in some region. This interpretation seems confirmed in appendix \ref{ap:old}. Finally, in appendix \ref{ap:noncomp}, we provide an analytical resolution of the eigenmode equation in the non-compact case, also corresponding to the behaviour close to a source: for an $O_p$, we conclude again on the presence of tachyons.

Neither these tachyons nor this negative region are however physical! As explained in the Introduction and section \ref{sec:backgd}, our supergravity description breaks down in the negative region and a proper description would require string theory. Our equations should not be trusted anymore in that region. Computing the spectrum, we should then find a way to fully ignore the effects of this region, in particular the tachyonic modes. To that end, we develop a procedure, presented in the following.

\subsection{Restricting the domain and reformulating the eigenmode equation}\label{sec:D}

\begin{figure}[h]
	\centering{
		\resizebox{80mm}{!}{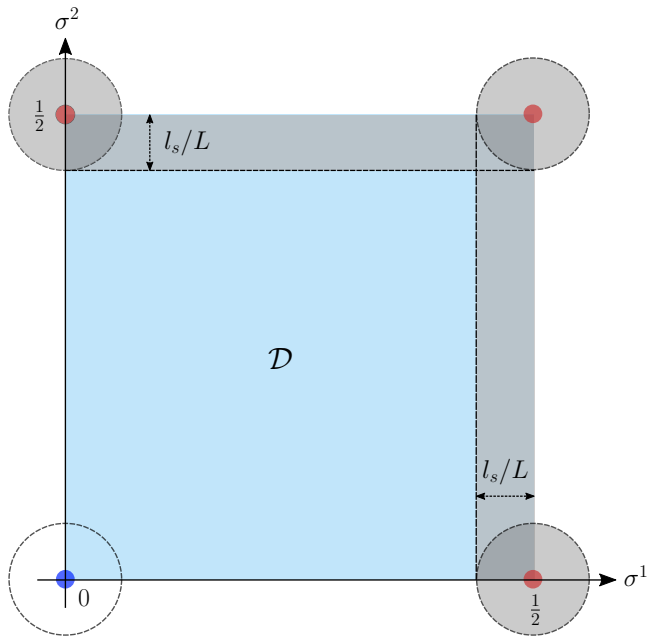}
\caption{Regions in a quarter of $\mathbb{T}^{d=2}$, the rest of the torus being obtained by axis symmetry (as in Figure \ref{fig:H3intro} and \ref{fig:H3cut}). The $O_p$ are placed at the four corners, and depicted by red and blue dots; the one at the origin (blue dot) is not problematic thanks to the 16 additional $D_p$ present there which give an excess $D_p$ charge. The restricted domain ${\cal D}$ is light blue, while the excluded region is gray. The negative region corresponding to prescription \eqref{H0prescr} is darker gray and bounded by circles of radius $l_s/L$ around the $O_p$. In general, these regions are actually not exactly circular as discussed in section \ref{sec:away}. We see that the excluded region is larger than what is a priori needed, leading to the adjusted prescription \eqref{H0prescradj}.}\label{fig:D}
	}
\end{figure}

We introduce a domain ${\cal D}$ where the warp factor is non-negative, and solve the eigenmode equation on ${\cal D}$ only. We recall that $\sigma^m \in \ (-\tfrac{1}{2},\tfrac{1}{2}]$ and that $l_s/L < 1/2$. We also recall that $O_p$, close to which $H<0$, are placed precisely at $\sigma^m = \frac{1}{2}$ as specified at the beginning of section \ref{sec:towardsspectrum}. We then define
\beq
\lambda= 1- 2 \frac{l_s}{L} \ ,\qquad {\cal D} = \left\{ {\bm \sigma} , \ |\sigma^m| \leq \frac{1}{2} - \frac{l_s}{L} = \frac{1}{2} \lambda \right\} \ .
\eeq
The region of interest is thus reduced by a factor $\lambda$, as in Figure  \ref{fig:grapheintro}, and we introduce an appropriate coordinate to span it
\beq
\tau^m = \lambda^{-1}\, \sigma^m \quad \Rightarrow \quad {\cal D} = \left\{ {\bm \tau} , \ |\tau^m| \leq \frac{1}{2}  \right\} \ .
\eeq
This is designed to guarantee $H \geq 0$ on ${\cal D}$, relative to prescription \eqref{H0prescr} that refers to a distance $|{\bm \sigma} - {\bm \sigma}_j|=\frac{ l_s}{L}$ from an $O_p$ source $j$. This domain actually excludes a little more than needed by this prescription, as depicted for $d=2$ in Figure \ref{fig:D}. We will then adjust the prescription.

The restricted domain ${\cal D}$ leads us to slightly modify the prescription \eqref{H0prescr}. We now consider points of ${\cal D}$ which are the closest to the $O_p$ sources (except the one at the origin, due to the additional $D_p$-branes there). In $d=2$, as illustrated in Figure \ref{fig:D}, these are the corners of the white rectangle. Those points are simply given by ${\bm \sigma} = \lambda {\bm \sigma}_j$ for each source $j$ located by ${\bm \sigma}_j$. Indeed, the proportion $\lambda$ is preserved thanks to the Intercept Theorem (or Thales's Theorem). We then need to compare the values of $H$ at these various points, and take the most negative one. This gives the adjusted prescription
\beq
H_0 = \max_j \, \left\{- \sum_i Q_i\, G(\lambda {\bm \sigma}_j - {\bm \sigma}_i) \right\} \ .\label{H0prescradj}
\eeq
It would match the previous prescription \eqref{H0prescr} for a vector of norm $|{\bm \sigma}_j|=\frac{1}{2}$: in that case one has for $j=i$ that $|\lambda {\bm \sigma}_i - {\bm \sigma}_i| = 2 \frac{l_s}{L} |{\bm \sigma}_i|$ = $\frac{l_s}{L}$ as in \eqref{H0prescr}. Some ${\bm \sigma}_j$ have however different norms. In any case, the adjusted prescription \eqref{H0prescradj} guarantees that $H \geq 0 $ on ${\cal D}$. This gives appropriate boundary conditions to $H$ to have a well-defined spectral problem on ${\cal D}$.

We then consider the warp factor restricted to ${\cal D}$ only, and introduce some rescaling for convenience
\beq
\tilde{H}({\bm \tau}) = \lambda^2 \, \frac{1}{g_s} \left( \frac{L}{l_s} \right)^{d-2} \, H({\bm \sigma})\big|_{{\cal D}} \ , \qquad \tilde{H}_0 = \lambda^2 \, \frac{1}{g_s} \left( \frac{L}{l_s} \right)^{d-2} \, H_0 \ . \label{tildeH}
\eeq
The rescaling removes in the non-constant part of $H$ all dependence on the physical parameters $L$, $l_s$ and $g_s$, except through the overall $\lambda$. We now solve the eigenmode equation \eqref{psieq} on ${\cal D}$ only, considering now $\psi_N ({\bm \tau})$ defined in that restricted domain. We rewrite that equation on ${\cal D}$ as
\beq
-\delta^{mn} \frac{\partial}{\partial\tau^m}\frac{\partial}{\partial\tau^n} \psi_N({\bm \tau})=(2\pi\, \mu_N)^2\, \tilde{H}({\bm \tau})\,  \psi_N({\bm \tau}) \ , \label{psieqD}
\eeq
introducing as in \cite{Andriot:2019hay} the convenient eigenvalues
\beq
\mu_N^2 =  M_N^2 \, L^2\, g_s \left( \frac{l_s}{L} \right)^{d-2} \ .\label{muN}
\eeq
What will matter are not the actual eigenvalues, but their comparison to the standard spectrum \eqref{st} in the absence of sources. We now define the relevant ratio to the standard spectrum
\beq
f_N = \frac{M_N}{M_N^{{\rm (st)}}} = \frac{\mu_N}{\mu_N^{{\rm (st)}}} \ ,\quad {\rm where} \ \mu_N^{{\rm (st)}} = \frac{N}{ \sqrt{\lambda^{-2} \tilde{H}_0}} \ . \label{fN}
\eeq
In the following, we will refer to this ratio and its difference to 1 as the deviation, with respect to the standard spectrum.\\

To solve the eigenmode equation \eqref{psieqD}, we finally need a further rewriting. The warp factor and the eigenfunctions are now continuous functions defined on ${\cal D}$ only. We can consider them as periodic on successive copies of ${\cal D}$, up to cuts at points at the boundaries. This allows us to develop them as Fourier series on ${\cal D}$. Equivalently, as square integrable functions on an interval, they can be developed on the basis of functions formed by integer Fourier modes. Writing them as Fourier series on ${\cal D}$ corresponds to completely ignoring the negative region, and is again a way to provide appropriate boundary conditions for $H$ and $\psi_N$ to have a well-defined spectral problem. We now drop the $N$, considering only one eigenmode, and get
\beq
\tilde{H}({\bm \tau}) = \sum_{{\bm m} \in \mathbb{Z}^d} d_{{\bm m}}\, {\rm e}^{2 \i \pi\, {\bm m} \cdot {\bm \tau}} \,, \qquad \psi({\bm \tau}) = \sum_{{\bm m} \in \mathbb{Z}^d} c_{{\bm m}}\, {\rm e}^{2 \i \pi\, {\bm m} \cdot {\bm \tau}} \ . \label{expansions}
\eeq
From \eqref{psieqD} or \eqref{expansions}, the zero mode with $\mu_0 = 0$ is given by $\psi_0$ being a constant (a continuous harmonic function on a compact space). From now on, we consider other eigenmodes and $\mu \neq 0$. For each of these modes, the eigenmode equation \eqref{psieqD} becomes thanks to \eqref{expansions} the tower of equations
\begin{equation}
		\label{master}
\frac{{\bm n}^2}{\mu^2} c_{{\bm n}} - \sum_{{\bm m} \in \mathbb{Z}^d} c_{{\bm m}} \, d_{{\bm n} - {\bm m}} = 0 \,, \qquad \forall {\bm n} \in \mathbb{Z}^d \ .
\end{equation}
The $c_{{\bm n}}$ are the variables, together with the unknown $\mu$, while the coefficients $d_{{\bm m}}$ are fixed by $\tilde{H}$. Let us determine the latter, before solving this reformulated eigenmode equation \eqref{master}.

One obtains the $d_{{\bm m}}$ as the following Fourier coefficients
\beq
d_{{\bm m}} = \lambda^2 \, \frac{1}{g_s} \left( \frac{L}{l_s} \right)^{d-2} \,  \int_{-\frac{1}{2}}^{\frac{1}{2}} \d^d {\bm \tau}\  {\rm e}^{-2 \i \pi\, {\bm m} \cdot {\bm \tau}} H(\lambda {\bm \tau}) \ .
\eeq
We further use the following Fourier series expression for $H$
\beq
H({\bm \sigma}) = H_0 -\frac{1}{(2\pi L)^{d-2}} \sum_{{\bm n}\in \mathbb{Z}^{d\, *}} e^{2\pi \i \, {\bm n}\cdot {\bm \sigma}} \times \sum_i Q_i  \frac{e^{-2\pi \i\, {\bm n}\cdot {\bm \sigma}_i}}{4\pi^2 {\bm n}^2} \ ,
\eeq
based on the Green's function \eqref{GFourier}. We recall that this last Fourier series is not absolutely convergent and requires regularization. However, as argued in \cite{Andriot:2019hay}, we will truncate this infinite sum, thus avoiding this issue. We will also verify numerically in section \ref{sec:num} the successful matching of the truncated Green's function and the proper expression \eqref{Gtheta}. Using the various definitions and sources positions, we rewrite the above as
\beq
H({\bm \sigma}) = g_s \left(\frac{l_s}{L} \right)^{d-2} \left( \lambda^{-2} \tilde{H}_0 +  \sum_{{\bm n}\in \mathbb{Z}^{d\, *}} e^{2\pi \i \, {\bm n}\cdot {\bm \sigma}} \times \frac{1}{4\pi^2 {\bm n}^2} \big( 16 - 2^{4-d} \sum_{O_p} e^{-2\pi \i\, {\bm n}\cdot {\bm \sigma}_i} \big) \right) \ .
\eeq
We deduce
\beq
\lambda^{-2} d_{{\bm m}} =  \delta_{{\bm m}, {\bm 0}}\, \lambda^{-2} \tilde{H}_0 + \sum_{{\bm n}\in \mathbb{Z}^{d\, *}} \frac{1}{4\pi^2 {\bm n}^2} \big( 16 - 2^{4-d} \sum_{O_p} e^{-2\pi \i\, {\bm n}\cdot {\bm \sigma}_i} \big)  \prod_{q=1}^{d} \frac{\sin(\pi (\lambda {\bm n} - {\bm m})^q )}{\pi (\lambda {\bm n} - {\bm m})^q} \ , \label{dm}
\eeq
where $q$ labels the $q$-component of the $d$-vectors. We note that for $\lambda \approx 1 $, i.e.~$L/l_s \gg 1$, the last product boils down to $\delta_{{\bm m}, {\bm n}}$, reproducing the result of \cite{Andriot:2019hay}. This is a consistency check since for $\lambda \approx 1 $, ${\cal D}$ matches the full $\mathbb{T}^d$, considered in that paper.

As a side remark, let us note that $d_{{\bm 0}}$ is not only $\tilde{H}_0 $, contrary to $H({\bm \sigma})$. This is because the average of the sum of Green's functions (or the varying part of $H$) is not vanishing over the restricted domain. More precisely, one verifies that it is vanishing for $\lambda \approx 1$ and ${\bm m} = {\bm 0}$ thanks to the cancelation of charges, but not otherwise. This is only an observation since we will not make use of this zero-mode $d_{{\bm 0}}$, which is also the average of $\tilde{H}$. In particular, the deviation of the spectrum with respect to the case of a ``constant warp factor'' is computed with the standard spectrum \eqref{st}, corresponding to the case without any source.

Having determined the $d_{{\bm m}}$, the problem now amounts to solving the tower of equations \eqref{master}, to obtain the eigenvalues $\mu$ and corresponding eigenfunctions in terms of their coefficients $c_{{\bm n}}$. More precisely, we are interested in the deviation $f_{N}$ \eqref{fN} between the eigenvalue for a (massive) state and the corresponding standard value, in the absence of sources. We now turn to the numerical method used to solve the tower of equations \eqref{master} and thus the eigenmode equation, determining this way the spectrum.

\subsection{Numerical method to determine the spectrum}\label{sec:num}

Determining the mass spectrum of Kaluza--Klein gravitational waves on our $D_p/O_p$ warped toroidal background amounts to solving the eigenmode equation, decomposed into a tower of equations \eqref{master}. To that end, we present in the following and in appendix \ref{ap:num} the numerical method used. Its starting point is similar to the one of \cite{Andriot:2019hay}, namely having a vanishing determinant, but we improve it on many levels to be detailed, allowing us to reach a more precise spectrum, with more eigenmodes and in more dimensions ($d=1,...,6$, whereas we stopped at $d=3$ in \cite{Andriot:2019hay}). That spectrum is given in section \ref{sec:spectrum}.\\

\begin{figure}[H]
	\begin{center}
		\includegraphics[width=0.6\textwidth]{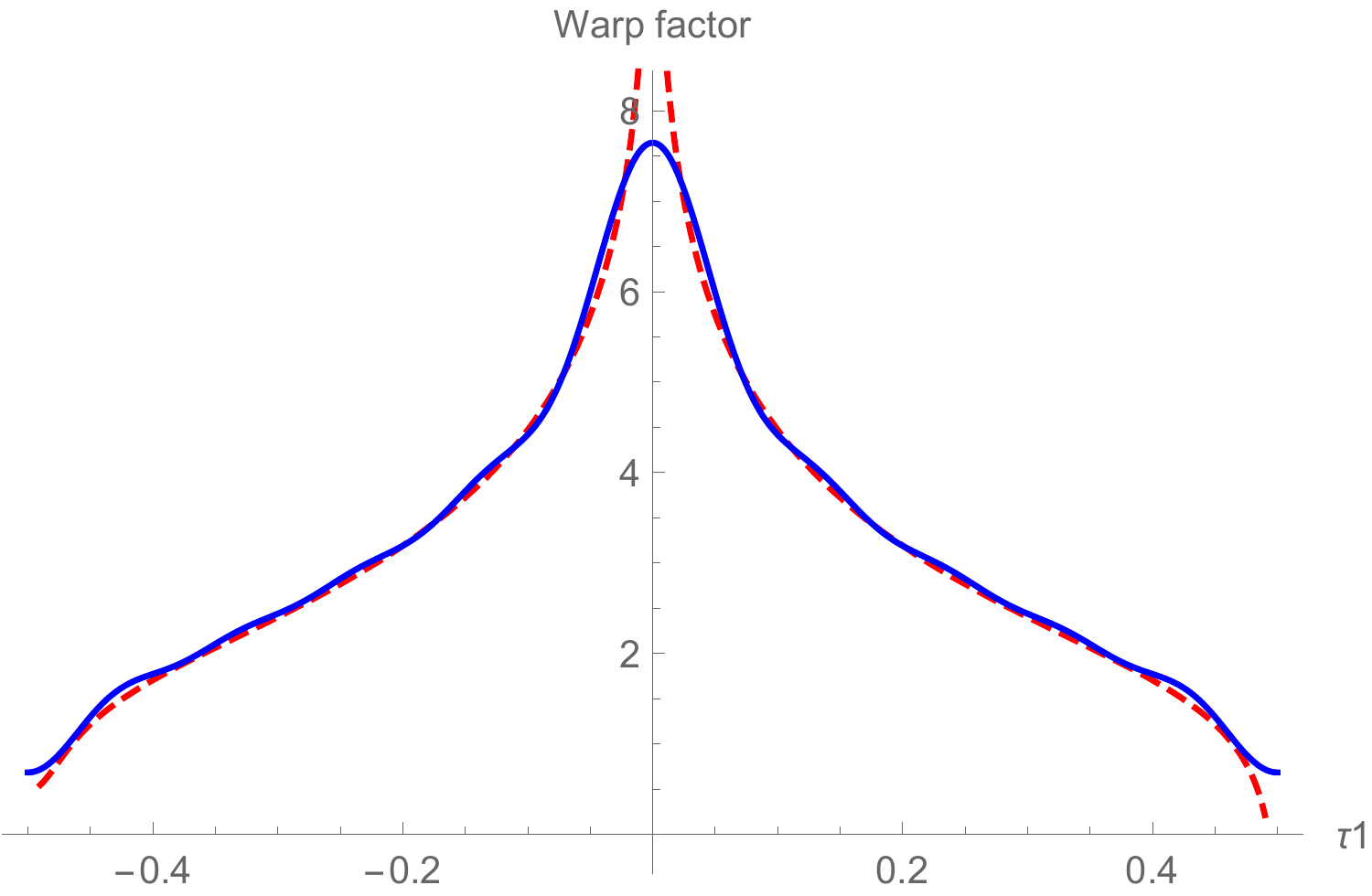}
		\caption{Comparison between the exact and the truncated warp factor. Here $d=2$ and the plots are along $\tau^1$, at fixed $\tau^2 = 0$. The red dashed line represents the exact $\tilde{H}$, while the blue line stands for the truncated $\tilde{H}$, with $r=10$ and $n=316$.} \label{fig:comparison}
	\end{center}
\end{figure}

\begin{figure}[h]
	\centering{
		\resizebox{80mm}{!}{\input{lattice.pdf_tex}}
		\caption{Example of the sample $\Gamma$ (in blue) of size $r$ for $d=2$, and the smaller sample $\tilde{\Gamma}$ (in red), sufficient to determine the spectrum. Here, $n=56$ and $\tilde{n} = 12$.}\label{fig:lattice}
	}
\end{figure}
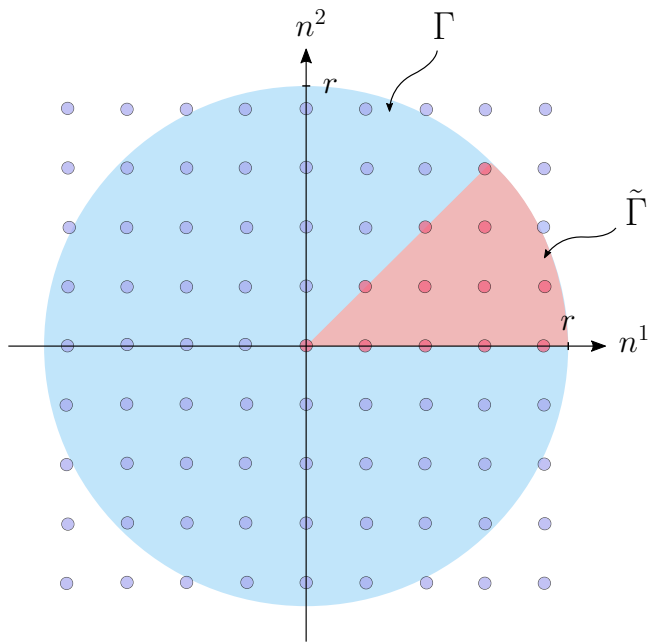

To deal with the tower of equations \eqref{master}, we start by imposing a truncation: we truncate the series \eqref{expansions} of the warp factor and the eigenfunction, keeping for each of them only a finite sample $\Gamma$ of the Fourier modes in momentum space, depicted in Figure \ref{fig:lattice}. This is done by retaining momenta whose norm is smaller than a value $r$. The size of the sample is denoted $n = \dim \Gamma$, and typically goes as $r^d$. As will be detailed, the larger $r$ (and $n$), the better the precision. We will be able to reach large $n$, and verify on that occasion that the truncated warp factor matches well its formal expression \eqref{H}, \eqref{Gtheta}, as depicted on Figure \ref{fig:comparison}.

We then establish a bijective map between the momenta ${\bm m}$ kept in $\Gamma$ and integers from $1$ to $n$ (see appendix \ref{ap:num} for more details). Improvements have been made w.r.t.~\cite{Andriot:2019hay} in establishing this sample and this map. The latter allows to write the coefficients $c_{{\bm m}}$ as a $n$-dimensional vector $\vec{c}$. The tower of equations \eqref{master} becomes a finite system, written in matrix form as
\begin{equation}
	\label{linear}
	\mathcal{O}(\mu) \cdot \vec{c} = 0 \ ,
\end{equation}
where $\mathcal{O}(\mu)$ is an $n\times n$ matrix, depending on the unknown $\mu$ and the $d_{{\bm m}}$ coefficients \eqref{dm}. Since $\psi$ is not identically zero by construction, and its $n$ first Fourier modes are assumed not to be all zero, $\vec{c}$ is not a null vector. So \eqref{linear} requires
\begin{equation}
	\label{det}
	\det \mathcal{O}(\mu) = 0 \ ,
\end{equation}
which can be further written as a polynomial equation. It can be solved numerically for a reasonable $n$ (in our case not larger than $\sim$ 80), since this operation has complexity\footnote{We measured that the time $t$ taken for such a computation obeys $\log t = C n + B_d$, where the constant $B_d$ is different from one dimension $d$ to another, while the constant $C$ seems to be the same for all $d$.} $\mathcal{O}({\rm e}^{C n})$. This does not allow for a high precision, especially for $d > 3$ where such a $n$ corresponds to a small radius $r$ (for $d=4$ one has $r \approx 2$), meaning that only the very first modes are not truncated. However, this first resolution still provides a good overview of the spectrum and its organization (degeneracy, approximate eigenvalues). In addition, for $d=1,2$, it already gives good estimates of the spectrum, detailed in section \ref{sec:spectrum}. In particular, this reveals that the largest observed deviation occurs for the lightest (massive) mode. In addition, the deviation seems to grow with $d$; this motivates us to access the spectrum for $d \geq 3$ with a satisfactory accuracy. As this requires a larger and larger number of points $n$, we need yet another rewriting of the problem to be solved. \\

To reduce the number of variables and equations to solve, and thus be able to reach a higher precision, we can make use of the symmetries of the source distribution, specified at the beginning of section \ref{sec:towardsspectrum}. While the $D_p$ are at the origin, the orientifold $O_p$ are, for each circle of $\mathbb{T}^d$, at the 2 fixed points at $\sigma^m=0$ or $\frac{1}{2}$. Considering the $O_p$ with mirrors (e.g.~at $\sigma^m=\pm \frac{1}{2}$), we see them placed for $d=2$ at the corners or on the edges of a square (see Figure \ref{fig:H3cut} and \ref{fig:D}), and generalizations thereof in higher dimensions $d$. This distribution is invariant under the exchange of the $O_p$: these permutations form the so-called hyperoctahedral group of degree $d$, that we denote $G$. For $d=2$, one has $G=D_4$ with $\dim D_4 = 8$. An interesting consequence is that for $d\geq 2$, $H$ and the lightest eigenmode $\psi_1$  inherit this symmetry\footnote{Let us note that for the special case $d=1$, the lightest mass is rather associated to an odd mode. For $d\geq 2$, the higher modes can be odd under certain transformations of $G$, and even under others.}. This gets translated in their Fourier coefficients, which are related to each other under transformations of $G$
\begin{equation}
	\label{Hsym}
	H(g \cdot \vs) = H(\vs) \quad \Longleftrightarrow \quad d_{g \cdot \vn} = d_{\vn} \,, \qquad \forall \vn \in \mathbb{Z}^d\,, \qquad \forall g \in G \,.
\end{equation}
\begin{equation}
	\label{psisym}
	\psi_1(g \cdot \vs) = \psi_1(\vs) \quad \Longleftrightarrow \quad c_{g \cdot \vn} = c_{\vn} \,, \qquad \forall \vn \in \mathbb{Z}^d\,, \qquad \forall g \in G \,.
\end{equation}
As noted previously, the highest deviation from the standard spectrum always occurs for the lightest mode, so we focus on the latter and its associated mass $\mu_1$ in the following, and assume the property \eqref{psisym}. This observation allows us to reduce drastically the number of independent Fourier coefficients to determine (for $d=6$, this number is reduced by a factor $\mathcal{O}(10^5)$), and correspondingly the number of independent equations \eqref{master}. We reach this way higher $r$ and $n$, i.e.~larger samples, necessary to get a reasonable precision in high dimensions $d$. This last improvement w.r.t.~\cite{Andriot:2019hay} was crucial to get interesting results on the spectrum for $d=4,5,6$.
The details of this technical simplification are given in appendix \ref{ap:num}. The upshot is that one can merely consider the points in the sample $\Gamma$ that are not equivalent under the action of $G$, i.e. the points
\begin{equation}
	\vm \in \tilde{\Gamma} = \frac{\Gamma}{G} \,, \qquad \tilde{n} = \dim \tilde{\Gamma} = \frac{\dim \Gamma}{2^d \, d!} \,.
\end{equation}
For instance, for $d=2$, one can consider the following representatives,
\begin{equation}
	\label{rep}
	\tilde{\Gamma} = \left\{\vm \in \Gamma, \; m^1 \geq 0 \; \text{and} \; 0 \leq m^2 \leq m^1  \right\} \,,
\end{equation}
as illustrated on Figure \ref{fig:lattice}.

Eventually, one has $\tilde{n}$ Fourier coefficients $d_{{\bm m}}$ to compute, and $\tilde{n}$ equations to solve
	\begin{equation}
		\label{syst}
		\frac{|\vn|^2}{\mu_1^2} c_{\vn} - \sum_{\vm \in \Gamma} c_{R(\vm)} \, d_{R(\vn - \vm)} \,  = 0 \,, \qquad \forall \vn \in \tilde{\Gamma} \,,
	\end{equation}
where $R(\vm)$ denotes the representative of $\vm$ in $\tilde{\Gamma}$. We refer to appendix \ref{ap:num} for its explicit implementation, and for the derivation of \eqref{syst}.\\

As explained above, we start by solving the determinant equation \eqref{det}. This provides a good overview of the spectrum, but also a good estimate of the lightest mass that we are interested in. In fact, we use this first value as a ``seed'' for our algorithm: we make use of minimization techniques that look for a solution to the system of equations \eqref{syst} close to the seed, this time with more Fourier modes (i.e. larger $r$ and $n$). More specifically, we run a \texttt{FindMinimum} in Mathematica, where the quantity to be minimized is the sum of the squares of the left hand sides in \eqref{syst} (see e.g.~appendix B of \cite{Andriot:2020wpp} for a more detailed description of this method, in a different context). The output provides a refined value for $\mu_1$ and the deviation $f_1$. We then repeat the same step, and use the refined value as a new seed, and with a larger $r$ (and $n$). We repeat this procedure until we reach a satisfactory precision on $\mu_1$ and $f_1$, i.e. when the latter converge towards fixed value, typically when the value does not vary of more than $1\%$ with respect to the previous step. This method allows to reach very large $n$, at most $n \sim 1.5 \cdot 10^6$ for $d=6$. This is a significant improvement with respect to the first attempts made in \cite{Andriot:2019hay}, that reached at most $n \sim 50$. It also allows us to reach $r \sim 10$ for every dimension $d$, and to rigorously compare the deviation obtained in each case.\\

Thanks to all these innovations, this method provides us with interesting, precise and new results on the spectrum of Kaluza--Klein gravitational waves on this warped toroidal background. We now turn to those.

\section{Spectrum}\label{sec:spectrum}

We present in this section the spectrum obtained for Kaluza--Klein gravitational waves on the warped toroidal background of section \ref{sec:backgd}. The $D_p/O_p$ sources generating the warping are placed as specified at the beginning of section \ref{sec:towardsspectrum}. The numerical method used to determine this spectrum is presented in section \ref{sec:num} and appendix \ref{ap:num}. While based on initial ideas of \cite{Andriot:2019hay}, it got improved on many levels allowing us to present here more precise results as well as new results, especially regarding the higher dimensions $d=4,5,6$. We also had to face the issue of the negative region, where $H<0$, that leads to tachyons in the spectrum as discussed in section \ref{sec:neg}, appendix \ref{ap:noncomp} and \ref{ap:old}. Thanks to a restriction of the domain on which relevant functions are defined, we could overcome this issue, as described in section \ref{sec:D}. The spectrum obtained prior to this restriction is given for completeness in appendix \ref{ap:old} and exhibits (unphysical) tachyonic modes, while the resulting tachyon-free spectrum is given in the following.

When giving the spectrum, we display the eigenvalues $\mu_N$ defined in \eqref{muN}, proportional to the Kaluza--Klein masses $M_N$. More importantly, we provide the value of $f_N = M_N/ M_N^{{\rm (st)}} = \mu_N / \mu_N^{{\rm (st)}} $ given in \eqref{fN}, the ratio to the standard mass or eigenvalue in absence of sources. The deviation $1 - f_N$ evaluates the difference from a standard toroidal spectrum, and thus the impact of a non-trivial warp factor. As recalled in \eqref{matchst} and several occurrences in section \ref{sec:towardsspectrum}, the standard spectrum is recovered at $L/l_s \gg 1$ (for $d\geq 2$). This is verified in the tables below with $f_N$ close to 1. We also recall that $L/l_s > 2$, so the largest deviations we will observe will happen at ``low'' $L/l_s \approx 10$. It is also for those values of $L/l_s$ that reaching a satisfactory precision is the most difficult.\\

We start by giving in tables \ref{tab:detd1}, \ref{tab:detd2} and \ref{tab:detd3} the spectrum obtained for $d=1,2,3$ using simply the determinant method described in section \ref{sec:num}. We present the first modes of the tower, specifying for each dimension $d$ the size $n$ of the sample considered and the corresponding radius $r$ (see section \ref{sec:num}). The corresponding precision is already good for $d=1,2$, but will be improved below for $d=3$. The modes of the standard spectrum are labeled by an integer $N$: at each such level, the modes obey a certain degeneracy. This degeneracy gets partially lifted when moving away from that standard spectrum by lowering $L/l_s$. We distinguish the modes whose eigenfunction is symmetric ($s$) or antisymmetric\footnote{Antisymmetric modes would be projected out by orientifolds, since the metric should be symmetric under the involution. We keep them here for completeness.} ($a$) under $\vs \rightarrow -\vs$, and indicate in parentheses for each of those their degeneracy, e.g. $s(2)$. The combinatorics explaining these degeneracies are discussed in appendix \ref{ap:old}. The modes are ordered in the tables according to their mass at high $L/l_s$; when lowering the latter, we note that this order can get modified.
\begin{table}[H]
	\begin{center}
		\begin{tabular}{|c||c||c|c||c|c|}
			\hline
			& $N$ & \multicolumn{2}{c||}{$1$} & \multicolumn{2}{c|}{$2$} \\
			\cline{2-6}
			$L/l_s$ & $a/s$ &$a$& $s$ & $a$ &$s$ \\
			\hline
			\hline
			& $\mu_N$ & 1.369 & 1.593 & 2.846 & 3.081  \\
			\cline{2-6}
			$10$ & $f_N$   & 1.061 & 1.234 &  1.102 & 1.193 \\
			\hline
			\hline
			& $\mu_N$ & 1.010 & 1.175 & 2.099 & 2.273  \\
			\cline{2-6}
			$10^2$ & $f_N$   & 0.9897 & 1.151 &  1.028 & 1.113 \\
			\hline
			\hline
			& $\mu_N$ & 0.9829 & 1.143 & 2.043 & 2.211  \\
			\cline{2-6}
			$10^3$ & $f_N$   & 0.9809 & 1.141 &  1.019 & 1.103 \\
			\hline
		\end{tabular}
		\caption{Spectrum of the first Kaluza--Klein modes for $d=1$, with eigenvalue $\mu_N$ and deviation $f_N$ from the standard spectrum, according to the value of $L/l_s$. Sample specifications: $r=20,\, n = 41$.}\label{tab:detd1}
	\end{center}
\end{table}

\begin{table}[H]
	\begin{center}
		\begin{tabular}{|c||c||c|c|c||c|c|c|}
			\hline
			& $N$ & \multicolumn{3}{c||}{$1$} & \multicolumn{3}{c|}{$\sqrt{2}$} \\
			\cline{2-8}
			$L/l_s$ & $a/s$ &$s(1)$& $a(2)$ & $s(1)$&$s(1)$ & $a(2)$ & $s(1)$ \\
			\hline
			\hline
			& $\mu_N$ & 1.028 & 1.069 & 1.311 & 1.571 & 1.742 & 1.870  \\
			\cline{2-8}
			$9$ & $f_N$   & 0.9046 & 0.9407 & 1.154 & 0.9775 & 1.084 & 1.164 \\
			\hline
			\hline
			& $\mu_N$ & 0.9916 & 1.034 & 1.269 & 1.517 & 1.679 & 1.794  \\
			\cline{2-8}
			$10$ & $f_N$   & 0.9115 & 0.9506 &  1.166 & 0.9863 & 1.092 & 1.166 \\
			\hline
			\hline
			& $\mu_N$ & 0.6105 & 0.6492 & 0.7251 & 0.9288 & 0.9576 & 1.005  \\
			\cline{2-8}
			$10^2$ & $f_N$   & 0.9295 & 0.9884 &  1.104 & 1.000 & 1.031 & 1.082 \\
			\hline
			\hline
			& $\mu_N$ & 0.4539 & 0.4735 & 0.4995 & 0.6719 & 0.6776 & 0.6925  \\
			\cline{2-8}
			$10^3$ & $f_N$   & 0.9519 & 0.9929 &  1.048 & 0.9964 & 1.005 & 1.027 \\
			\hline
		\end{tabular}
		\caption{Spectrum of the first Kaluza--Klein modes for $d=2$, with eigenvalue $\mu_N$ and deviation $f_N$ from the standard spectrum, according to the value of $L/l_s$. Sample specifications: $r=4.5,\, n = 69$.}\label{tab:detd2}
	\end{center}
\end{table}

\begin{figure}[H]
	\begin{center}
		\includegraphics[width=0.6\textwidth]{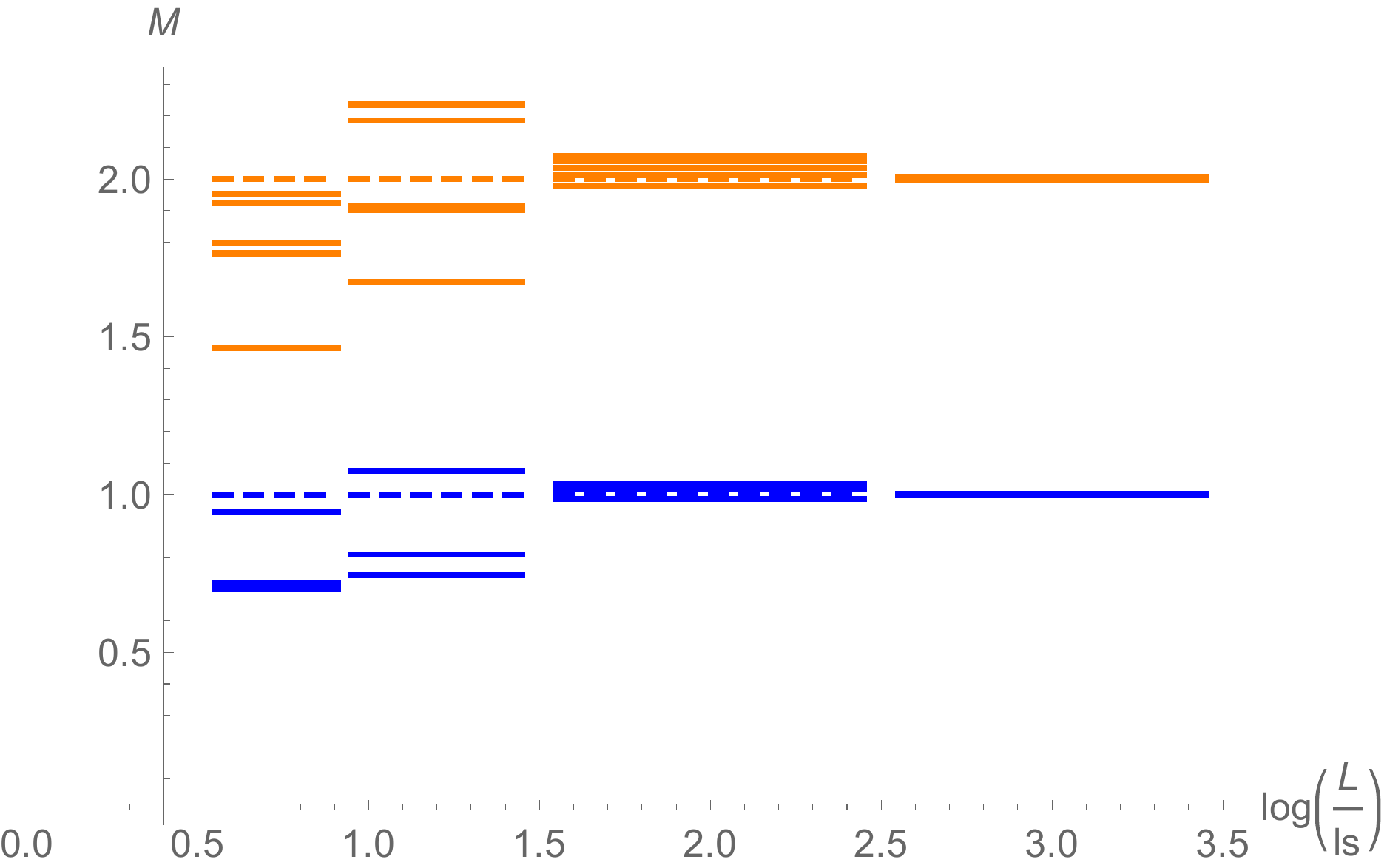}
		\caption{Mass spectrum for $d=3$ for the first two levels $N$, normalised to 1 and 2 for readability, in terms of $\log{\frac{L}{l_s}}$. We take for $\tfrac{L}{l_s}$ the four values of table \ref{tab:detd3}. The standard spectrum is represented by the dashed lines, and the spectrum obtained here with non-trivial warp factor by plain lines.} \label{fig:spectred3}
	\end{center}
\end{figure}

\begin{table}[H]
	\begin{center}
		\begin{tabular}{|c||c||c|c|c||c|c|c|c|c|}
			\hline
			& $N$ & \multicolumn{3}{c||}{$1$} & \multicolumn{5}{c|}{$\sqrt{2}$} \\
			\cline{2-10}
			$L/l_s$ & $a/s$ &$s(1)$& $a(3)$ & $s(2)$&$s(1)$ & $a(3)$ & $s(3)$  & $a(3)$& $s(2)$ \\
			\hline
			\hline
			& $\mu_N$ & 0.8353 & 0.8553 & 1.124 & 1.513 & 1.234 & 1.487 & 1.620 & 1.645 ?  \\
			\cline{2-10}
			$7$ & $f_N$   & 0.7012 & 0.718 & 0.9437 & 0.898 & 0.7322 & 0.8825 & 0.9618 & 0.9763 ? \\
			\hline
			\hline
			& $\mu_N$ & 0.7356 & 0.8012 & 1.063 & 1.330 & 1.172 & 1.339 & 1.528 & 1.563 ?  \\
			\cline{2-10}
			$10$ & $f_N$   & 0.744 & 0.8103 & 1.075 & 0.9509 & 0.8378 & 0.9579 & 1.093 & 1.118 ? \\
			\hline
			\hline
			& $\mu_N$ & 0.2534 & 0.2615 & 0.2655 & 0.3593 & 0.3658 & 0.3700 & 0.3739 & 0.3771  \\
			\cline{2-10}
			$10^2$ & $f_N$   & 0.9859 & 1.017 &  1.033 & 0.9883 & 1.006 & 1.018 & 1.029 & 1.037 \\
			\hline
			\hline
			& $\mu_N$ & 0.07941 & 0.07962 & 0.07972 & 0.1122 & 0.1125 & 0.1126 & 0.1127 & 0.1128 \\
			\cline{2-10}
			$10^3$ & $f_N$   & 0.9994 & 1.002 & 1.003 & 0.9986 & 1.001 & 1.002 & 1.003 & 1.004 \\
			\hline
		\end{tabular}
		\caption{Spectrum of the first Kaluza--Klein modes for $d=3$, with eigenvalue $\mu_N$ and deviation $f_N$ from the standard spectrum, according to the value of $L/l_s$. The question mark in the last entries indicates that the identification of the modes as being part of the same multiplet as those at higher $L/l_s$ is not certain. Sample specifications: $r=2.6,\, n = 81$.}\label{tab:detd3}
	\end{center}
\end{table}
The spectrum for $d=3$ is illustrated in Figure \ref{fig:spectred3} for the first two $N$. We see that the modes are degenerate to the standard spectrum values for large $L/l_s$, and the degeneracy gets lifted when lowering $L/l_s$, leading to the strongest deviations.

The values obtained in tables \ref{tab:detd1}-\ref{tab:detd3} should be compared to those of appendix \ref{ap:old} and to \cite{Andriot:2019hay}, where we did not restrict to the domain ${\cal D}$ that excludes the negative region, and took a constant value of $H_0$ instead of the prescribed one as here. The spectrum gets here corrected, and a crucial difference is the absence of tachyon for $d=2$ and $d=3$, when probing the same low values of $L/l_s$ as in appendix \ref{ap:old}. Regarding the deviation $f_N$, which is the interesting physical effect of the warp factor, we reach at best $f_N \approx 0.70$, i.e.~$30\%$. For $d\geq 2$, the smallest $f_N$ values are obtained for the lowest $L/l_s$, the lightest mode and the highest $d$. This motivates us in the following to further restrict the study to only the first mode with $\mu_1$ and investigate all $d$ up to 6. The more advanced numerical techniques presented in section \ref{sec:num} will allow us to do so, while reaching a better precision thanks to a larger sample.

Before this, let us briefly comment on $d=1$. This case is special, mostly because there is no divergence at the sources. As a consequence, one can find a finite constant $H_0$ which makes $H>0$ everywhere, as done around Figure 3 in \cite{Andriot:2019hay}. Here we still use the restricted domain and adjusted prescription \eqref{H0prescradj} for consistency, leading to a variation of the spectrum in $L/l_s$ (contrary to appendix \ref{ap:old} and \cite{Andriot:2019hay}). Despite this variation, it is not true for $d=1$ that the spectrum matches the standard one at large $L/l_s$ as discussed in \cite{Andriot:2019hay}. So $f_N$ does not particularly go to $1$ in that limit. Another observed specificity of $d=1$ is that the first massive mode is antisymmetric, contrary to higher $d$.\\

We turn to studying the lightest massive mode, whose deviation is observed to be the strongest. We give in tables \ref{tab:d1}-\ref{tab:d6} the values of $\mu_1$ and $1-f_1$ for $d=1, ..., 6$. We note that $1-f_1$ is not always positive, which means that the deviation from the standard spectrum can either increase or decrease the mass. We also give the constant $g_s^{-1} H_0$ obtained from the adjusted prescription \eqref{H0prescradj}. As mentioned above, its value for $d=1$ is special and varies strongly with $L/l_s$, but it is more regular for the other $d$. The precision reached is higher than above (in the sense of reducing the error), thanks to the additional steps described in section \ref{sec:num} allowing us to use a larger sample. The results are however only obtained with two significant digits.

\begin{table}[H]
	\begin{center}
		\begin{tabular}{|c||c|c|c|c|}
			\hline
     & & & & \\[-8pt]
			$L/l_s$ & 10 & $10^2$ & $10^3$ & $10^4$  \\[3pt]
			\hline
			\hline
      & & & & \\[-8pt]
			$g_s^{-1} H_0$ & 6.0 & 96 & $10^3$ & $10^4$  \\[3pt]
			\hhline{-||----}
      & & & & \\[-8pt]
			$\mu_1$ & 1.4 & 1.0 & 0.98 & 0.98  \\[3pt]
			\hhline{-||----}
      & & & & \\[-8pt]
			$1-f_1$ & $-0.061$ & 0.010 & 0.020 & 0.020  \\[3pt]
			\hline
		\end{tabular}
		\caption{First non-zero eigenvalue $\mu_1$ and deviation $f_1$ from the standard spectrum for $d=1$, together with the constant $g_s^{-1} H_0$ obtained from the adjusted prescription \eqref{H0prescradj}.}\label{tab:d1}
	\end{center}
\end{table}

\begin{table}[H]
	\begin{center}
		\begin{tabular}{|c||c|c|c|c|}
			\hline
     & & & & \\[-8pt]
			 $L/l_s$ & 10 & $10^2$ & $10^3$ & $10^4$  \\[3pt]
			\hline
			\hline
     & & & & \\[-8pt]
			$g_s^{-1} H_0$ & 0.84 & 2.3 & 3.8 & 5.2  \\[3pt]
			\hhline{-||----}
     & & & & \\[-8pt]
$\mu_1$ & 0.99 & 0.61 & 0.49 & 0.42  \\[3pt]
\hhline{-||----}
     & & & & \\[-8pt]
$1-f_1$ & 0.091 & 0.071 & 0.054 & 0.041  \\[3pt]
\hline
		\end{tabular}
		\caption{First non-zero eigenvalue $\mu_1$ and deviation $f_1$ from the standard spectrum for $d=2$, together with the constant $g_s^{-1} H_0$ obtained from the adjusted prescription \eqref{H0prescradj}.}\label{tab:d2}
	\end{center}
\end{table}

\begin{table}[H]
	\begin{center}
		\begin{tabular}{|c||c|c|c|c|}
			\hline
     & & & & \\[-8pt]
			$L/l_s$ & 10 & $10^2$ & $10^3$ & $10^4$  \\[3pt]
			\hline
			\hline
     & & & & \\[-8pt]
			$g_s^{-1} H_0$ & 0.1 & 0.15 & 0.16 & 0.16  \\[3pt]
			\hhline{-||----}
     & & & & \\[-8pt]
			$\mu_1$ & 0.65 & 0.25 & 0.079 & 0.025  \\[3pt]
			\hhline{-||----}
     & & & & \\[-8pt]
			$1-f_1$ & 0.34 & 0.015 & $6.4 \cdot 10^{-4}$ & $5.6 \cdot 10^{-5}$  \\[3pt]
			\hline
		\end{tabular}
		\caption{First non-zero eigenvalue $\mu_1$ and deviation $f_1$ from the standard spectrum for $d=3$, together with the constant $g_s^{-1} H_0$ obtained from the adjusted prescription \eqref{H0prescradj}.}\label{tab:d3}
	\end{center}
\end{table}

\begin{table}[H]
	\begin{center}
		\begin{tabular}{|c||c|c|c|c|}
			\hline
     & & & & \\[-8pt]
			$L/l_s$ & 10 & $10^2$ & $10^3$ & $10^4$  \\[3pt]
			\hline
			\hline
     & & & & \\[-8pt]
			$g_s^{-1} H_0$ & 0.012 & 0.025 & 0.025 & 0.025  \\[3pt]
			\hhline{-||----}
     & & & & \\[-8pt]
			$\mu_1$ & 0.35 & 0.064 & $6.3 \cdot 10^{-3}$  & $6.3 \cdot 10^{-4}$  \\[3pt]
			\hhline{-||----}
     & & & & \\[-8pt]
			$1-f_1$ & 0.62 & $-0.018$ & $-2.0 \cdot 10^{-3}$ & $-2.0 \cdot 10^{-4}$  \\[3pt]
			\hline
		\end{tabular}
		\caption{First non-zero eigenvalue $\mu_1$ and deviation $f_1$ from the standard spectrum for $d=4$, together with the constant $g_s^{-1} H_0$ obtained from the adjusted prescription \eqref{H0prescradj}.}\label{tab:d4}
	\end{center}
\end{table}

\begin{table}[H]
	\begin{center}
		\begin{tabular}{|c||c|c|c|c|}
			\hline
     & & & & \\[-8pt]
			$L/l_s$ & 10 & $10^2$ & $10^3$ & $10^4$  \\[3pt]
			\hline
			\hline
     & & & & \\[-8pt]
			$g_s^{-1} H_0$ & $3.9 \cdot 10^{-3}$ & $6.3 \cdot 10^{-3}$ & $6.3 \cdot 10^{-3}$ & $6.3 \cdot 10^{-3}$  \\[3pt]
			\hhline{-||----}
     & & & & \\[-8pt]
			$\mu_1$ & $<0.17$ & 0.013 & $4.0 \cdot 10^{-4}$  & $1.3 \cdot 10^{-5}$  \\[3pt]
			\hhline{-||----}
     & & & & \\[-8pt]
			$1-f_1$ & $ >0.65$ & $-0.02$ & $-2.0 \cdot 10^{-3}$ & $-2.0 \cdot 10^{-4}$  \\[3pt]
			\hline
		\end{tabular}
		\caption{First non-zero eigenvalue $\mu_1$ and deviation $f_1$ from the standard spectrum for $d=5$, together with the constant $g_s^{-1} H_0$ obtained from the adjusted prescription \eqref{H0prescradj}.}\label{tab:d5}
	\end{center}
\end{table}

\begin{table}[H]
	\begin{center}
		\begin{tabular}{|c||c|c|c|c|}
			\hline
     & & & & \\[-8pt]
			$L/l_s$ & 10 & $10^2$ & $10^3$ & $10^4$  \\[3pt]
			\hline
			\hline
     & & & & \\[-8pt]
			$g_s^{-1} H_0$ & $1.6 \cdot 10^{-3}$ & $2.0 \cdot 10^{-3}$ & $2.0 \cdot 10^{-3}$ & $2.0 \cdot 10^{-3}$  \\[3pt]
			\hhline{-||----}
     & & & & \\[-8pt]
			$\mu_1$ & $<0.078$ & $2.3 \cdot 10^{-3}$ & $2.2 \cdot 10^{-5}$  & $2.2 \cdot 10^{-7}$  \\[3pt]
			\hhline{-||----}
     & & & & \\[-8pt]
			$1-f_1$ & $>0.69$ & $-0.02$ & $-2.0 \cdot 10^{-3}$ & $-2.0 \cdot 10^{-4}$  \\[3pt]
			\hline
		\end{tabular}
		\caption{First non-zero eigenvalue $\mu_1$ and deviation $f_1$ from the standard spectrum for $d=6$, together with the constant $g_s^{-1} H_0$ obtained from the adjusted prescription \eqref{H0prescradj}.}\label{tab:d6}
	\end{center}
\end{table}

\begin{figure}[H]
	\begin{center}
		\includegraphics[width=0.6\textwidth]{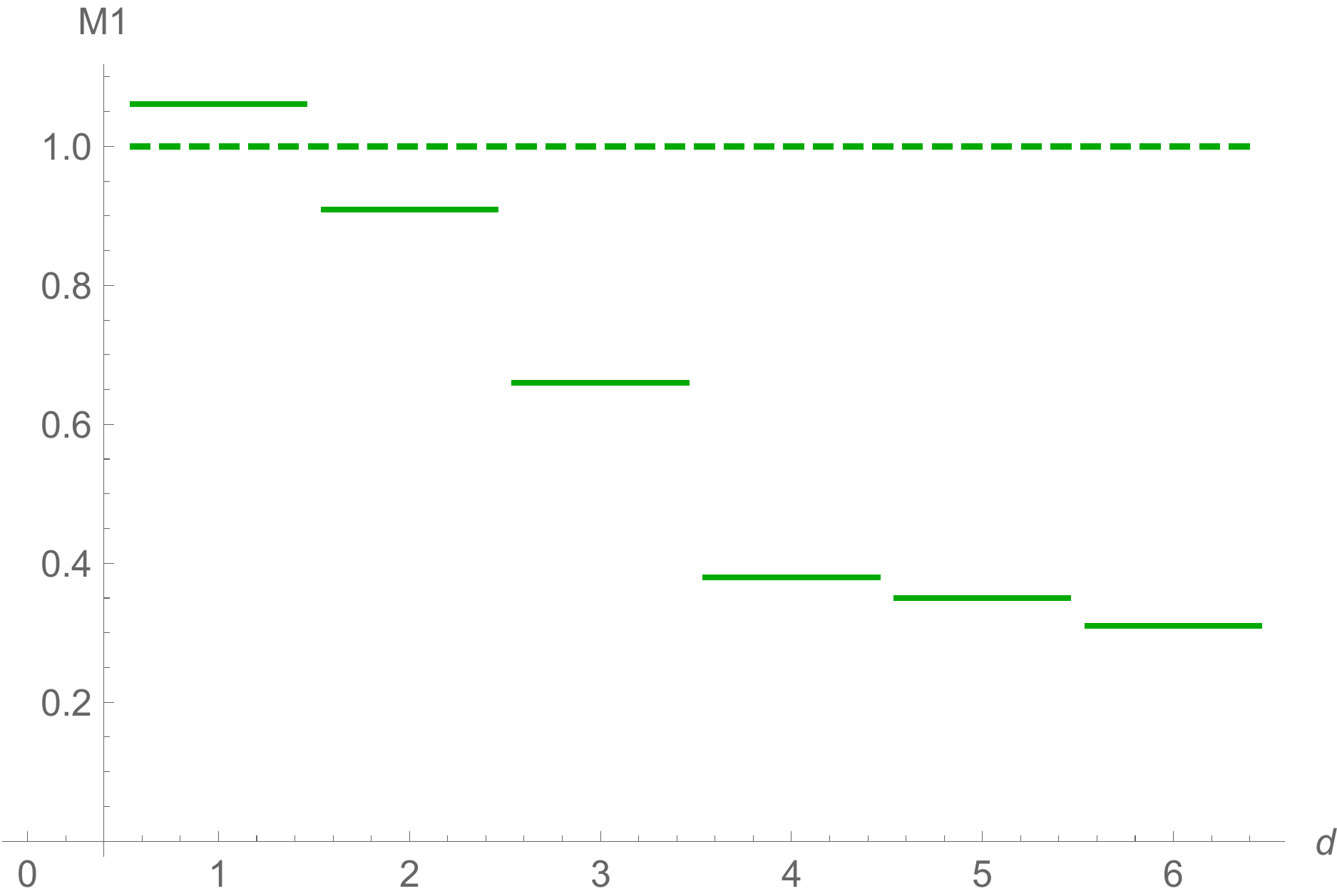}
		\caption{First non-zero mass $M_1$ for $d=1, ..., 6$ (plain lines), compared to that of the standard spectrum normalised to 1 (dashed line), for $L/l_s =10$.} \label{fig:spectredalld}
	\end{center}
\end{figure}

The largest deviation $|1-f_1|$ is obtained for the lowest $L/l_s$ and highest dimension: it is of $69\%$ for $d=6$. This strong deviation is one of the main results of this analysis. We illustrate the evolution of the first non-zero mass in terms of the dimension in Figure \ref{fig:spectredalld}, compared to the standard spectrum. We finally make side comments on the values of $g_s^{-1} H_0$: those match the ones obtained analytically in \eqref{constantH0} at large $L/l_s$ for $d\geq 3$, and slowly vary from there. This however does not hold for $d=2$: the constant obtained by the prescription, while obeying the behaviour of \eqref{constantH0} in $L/l_s$, seems to differ by a constant shift of approximately $0.6$. This could correspond to the next order after the leading behaviour of \eqref{behiaviourG}.

\section{Summary}\label{sec:ccl}

Kaluza--Klein gravitational waves, most likely of primordial origin, would provide a very specific signature of extra dimensions, in the form of a specific discrete spectrum. Motivated by this possibility, we determined in this work such a spectrum in the case where the extra dimensions include a warped torus $\mathbb{T}^d$.\footnote{Considering other compact spaces would be very interesting, especially those appearing in different string compactifications. To that end, the Laplacian spectrum on e.g.~a nilmanifold \cite{Andriot:2018tmb} or a Calabi-Yau manifold \cite{Ashmore:2020ujw} would be useful.} While the Kaluza--Klein spectrum on a torus is well-known (we refer to it as the standard spectrum), we aimed here at measuring the effect of a non-trivial warp factor $H$, by evaluating the deviation of the resulting spectrum with respect to the standard one. We recall that a warp factor is a generic ingredient in BSM models as well as string compactifications, that captures the back-reaction of $D_p$-branes, as well as here orientifold $O_p$-planes. Characterising its impact on the spectrum is thus important. We considered the warp factor that would appear in common toroidal string compactifications, determined in \cite{Andriot:2019hay}, which differs from other options in the literature such as that of Randall-Sundrum models \cite{Randall:1999ee, Randall:1999vf}, providing interesting novelties. We first tackled this problem in \cite{Andriot:2019hay} and reviewed the relevant material in section \ref{sec:backgd} and \ref{sec:gwsp}. New information on the profile of the Green's function and the warp factor away from the sources is discussed in section \ref{sec:away}. We then made here important technical progress that eventually allowed a more precise and extended determination of the spectrum. The results are presented in section \ref{sec:spectrum}. We first gave in tables \ref{tab:detd1}-\ref{tab:detd3} the spectrum for the first massive Kaluza--Klein gravitational waves for $d=1,2,3$, according to the ratio $L/l_s$ of the radius of $\mathbb{T}^d$ to the string (or fundamental) length. For large $L/l_s$, one recovers the standard spectrum, while for small ratios, $2 < L/l_s < 10$, the deviation is the most important. We illustrated these results, as well as the corresponding lift of degeneracies of the Kaluza--Klein modes, in Figure \ref{fig:spectred3}. We then focused on the lightest (massive) mode, for which the deviation is observed to be the highest, and we determined it for $d=1,...,6$ in tables \ref{tab:d1}-\ref{tab:d6} according to $L/l_s$. We illustrated the results in Figure \ref{fig:spectredalld}: the highest deviation is obtained for $d=6$ at low $L/l_s$, and amounts to $69\%$. The Kaluza--Klein mass is thus significantly lowered because of the warp factor.

To reach these results, we had to overcome an important physical challenge: the negative region, depicted in Figure \ref{fig:grapheintro}, where $H<0$. As discussed in the Introduction, the warp factor, and this negative region in particular, are currently at the center of many different investigations in string theory, and the present work may provide insights to these other related topics. We recalled there that $H <0$ close to orientifolds in the supergravity description. This is however unphysical, so this description is expected to break down, and be resolved by extra stringy physics. For this reason, the prescription of \cite{Andriot:2019hay} was to place the ``horizon'' where $H=0$ at a string length from $O_p$. Thanks to the extra numerical precision gained here, we however realised that the negative region would generate tachyonic modes in the spectrum (see appendix \ref{ap:old}). We argue in section \ref{sec:neg} and appendix \ref{ap:noncomp} why this should happen in general. We excluded completely this negative, unphysical region from our analysis, and solved the eigenmode equation, that determines the spectrum, on a restricted domain ${\cal D}$ where $H\geq 0$. We described this restriction in detail in section \ref{sec:D} and illustrated the domain ${\cal D}$ in Figure \ref{fig:D}. This resolution, as well as several important numerical improvements presented in section \ref{sec:num} and appendix \ref{ap:num}, provided us with the tachyon-free, precise spectrum of Kaluza--Klein gravitational waves on a warped toroidal background described above.

\vspace{0.4in}

\subsection*{Acknowledgements}

D.~A.~acknowledges support from the Austrian Science Fund (FWF): project number M2247-N27.

\newpage

\begin{appendix}

\section{Tachyon in a non-compact case}\label{ap:noncomp}

In this appendix, we show analytically in a specific example the presence of tachyonic eigenmodes, due to the negative region close to an orientifold $O_p$, as discussed in section \ref{sec:neg}; for completeness we also study the case of a $D_p$-brane. We consider the situation where the approximate behaviour \eqref{behiaviourG} close to the source can be used, which also corresponds to the standard non-compact Green's function and warp factor. The differential problem to solve is the eigenvalue equation \eqref{psieq}, which takes the form
\eq{\label{1}
\Delta \psi({\bm r})+\tilde{M}^2\, H({\bm r})\, \psi({\bm r})=0
~,}
where $H$ is the warp factor, ${\bm r}$ parameterizes the position in the transverse torus $\mathbb{T}^d$, of Laplacian $\Delta$. We seek to determine the allowed eigenvalues $\tilde{M}^2$, especially their sign. We focus in the following on $d=3$ for simplicity.

\subsection{Orientifold source and negative region}\label{ap:Op}

Let us consider the problem in the vicinity of the origin, ${\bm r}={\bm 0}$ where an orientifold $O_6$ is located. The warp factor then takes the approximate form
\eq{\label{2}
H\simeq H_1-\frac{C}{r} ~,}
where $r=|{\bm r}|$, and $H_1$, $C$ are positive constants. We consider at first the problem in the region $0\leq r\leq C/H_1$, where $H$ is negative. We take the eigenfunction $\psi$ to only depend on the radial coordinate $r$. The differential equation \eqref{1} then reduces to
\eq{\label{3}
\left[r^2 \frac{\d^2}{\d r^2}+2r\frac{\d}{\d r}+\lambda(C r-H_1 r^2)\right]\psi(r)=0
~,}
where we have set $\lambda=-\tilde{M}^2$. By rescaling $\lambda$, $r$, we may henceforth set $H_1 = C=1$, without loss of generality.
Two linearly independent solutions
\eq{
e^{-\sqrt{\lambda} ~\!r}\ {}_1F_1(1-\tfrac{1}{2}\sqrt{\lambda} ~\!;2~\!;2\sqrt{\lambda} ~\!r)~;~~~
e^{-\sqrt{\lambda} ~\!r}\ U(1-\tfrac{1}{2}\sqrt{\lambda} ~\!,2~\!,2\sqrt{\lambda} ~\!r)
~,
}
can be  given in terms of the confluent hypergeometric functions of the first and second kind, ${}_1F_1$, $U$, also known as Kummer's and Tricomi's function respectively. For $-b\notin \mathbb{N}$ as here, ${}_1F_1(a ~\!;b~\!;z)$ is an entire function of both $a, z\in\mathbb{C}$. Tricomi's function can be defined in terms of ${}_1F_1(a~\!;b~\!;z)$.\footnote{This can be seen e.g.~in equation (7) of p.257 of \cite{Bateman}. That equation can be extended to all $c\in\mathbb{Z}$ by continuity.} For our purposes it suffices to note that $U(a,2,z)$ asymptotes $\tfrac{1}{\Gamma(a)z}$, as $z\rightarrow 0$.

\begin{figure}[H]
\begin{center}
\begin{subfigure}[H]{0.4\textwidth}
\includegraphics[width=\textwidth]{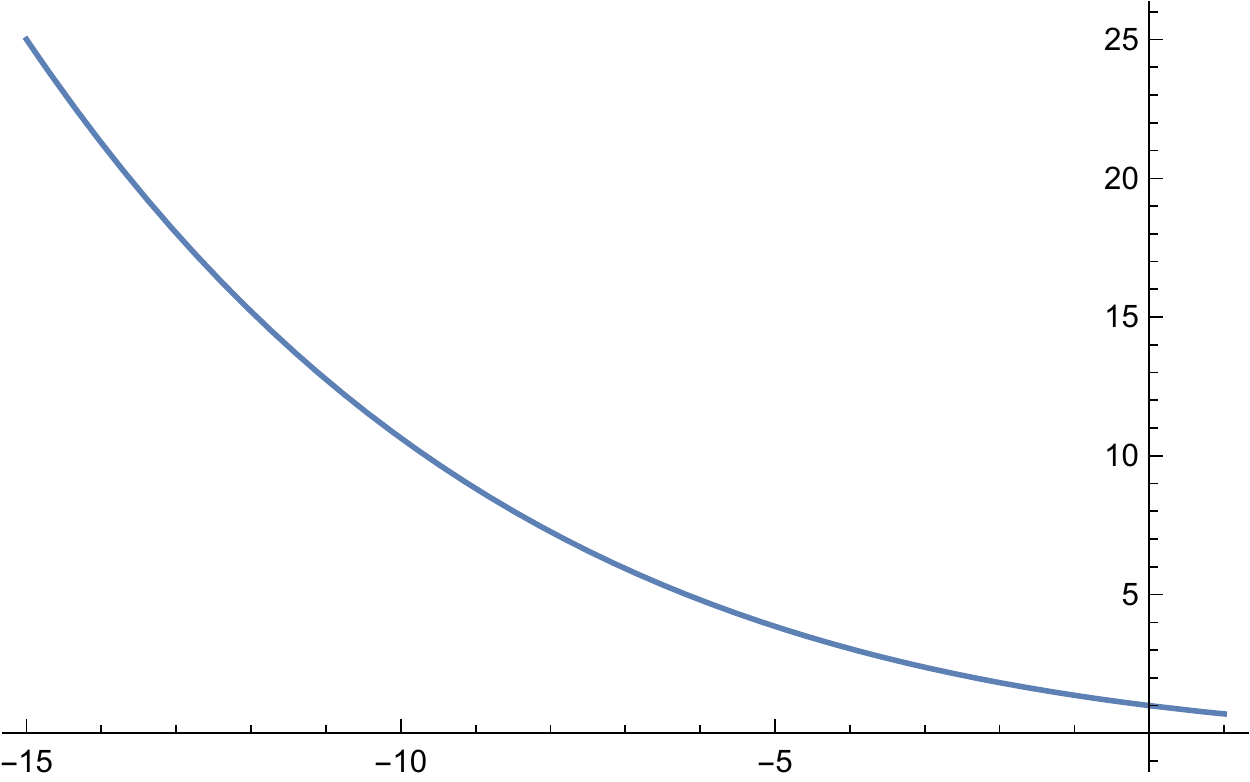}\caption{$\lambda\leq 1$}\label{fig:1a}
\end{subfigure}
\qquad \qquad
\begin{subfigure}[H]{0.4\textwidth}
\includegraphics[width=\textwidth]{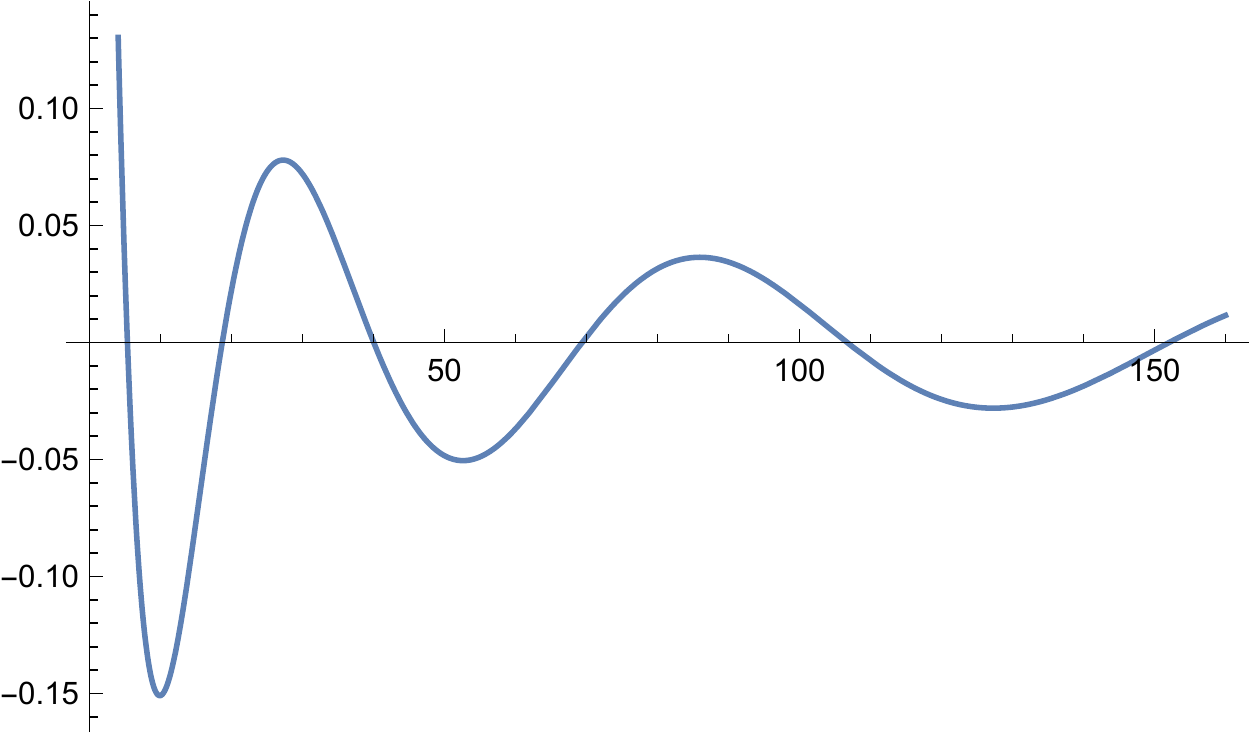}\caption{ $\lambda\geq 1$}\label{fig:1b}
\end{subfigure}
\caption{Graph of $f(\lambda)=e^{-\sqrt{\lambda} }\ {}_1F_1(1-\tfrac{1}{2}\sqrt{\lambda} ~\!;2~\!;2\sqrt{\lambda} )$  as a function of $\lambda$. At $\lambda=0$ we have $f=1$. For a better illustration, the graph is separated in two pieces, cutting at $\lambda=1$, where the function is continuous. Importantly, all the zeros of $f$ are located at positive $\lambda$.
}\label{fig:1}
\end{center}
\end{figure}

The first of these solutions is real-valued for any $\lambda\in\mathbb{R}$, as can be seen from the defining series expansion of Kummer's function (equation (1) of p.248 of \cite{Bateman}); in particular, $\lambda$ is allowed to be negative (despite the above notation $\sqrt{\lambda}$). To obtain a discrete spectrum for $\lambda$, we may impose ``separated'' boundary conditions. As an example, we pick
\eq{\label{sbc5}
\psi(0)= \psi(r_0)=0
~,}
for some fixed radial distance $r_0>0$. The first boundary condition then implies $\psi(r)\propto {}_1F_1$, since $U$ diverges at $r=0$. For fixed $r_0$, the second boundary condition only has solutions for a discrete spectrum of values for $\lambda$. For example, let us consider $r_0=1$, which is the location of the ``horizon'' where the warp factor vanishes. Figure \ref{fig:1} depicts $f(\lambda)=e^{-\sqrt{\lambda} ~\!r_0}\ {}_1F_1(1-\tfrac{1}{2}\sqrt{\lambda} ~\!;2~\!;2\sqrt{\lambda} ~\!r_0)$  as a function of $\lambda$, for $r_0 =1$. The spectrum of $\lambda$ is then given by the zeros of the function $f$. We note that there are no zeros for negative $\lambda$, i.e.~the spectrum of $\tilde{M}^2$ is strictly negative.

These conclusions remain unchanged if we impose boundary conditions with $0<r_0<1$, so that the warp factor is negative over the whole domain of definition $[0,r_0]$ of the differential problem. However, the behaviour changes if we impose boundary conditions with $r_0>1$, in which case the warp factor does not have definite sign in the domain $[0,r_0]$: there are then zeros for both negative and positive $\lambda$. This means that the spectrum is not bounded, neither above nor below. We conclude that this example exhibits in any case tachyonic eigenmodes, i.e.~$\tilde{M}^2<0$, as long we probe the region where $H<0$.

\subsection{$D_p$-brane source}\label{sec:3}

\begin{figure}[H]
\begin{center}
\begin{subfigure}[H]{0.4\textwidth}
\includegraphics[width=\textwidth]{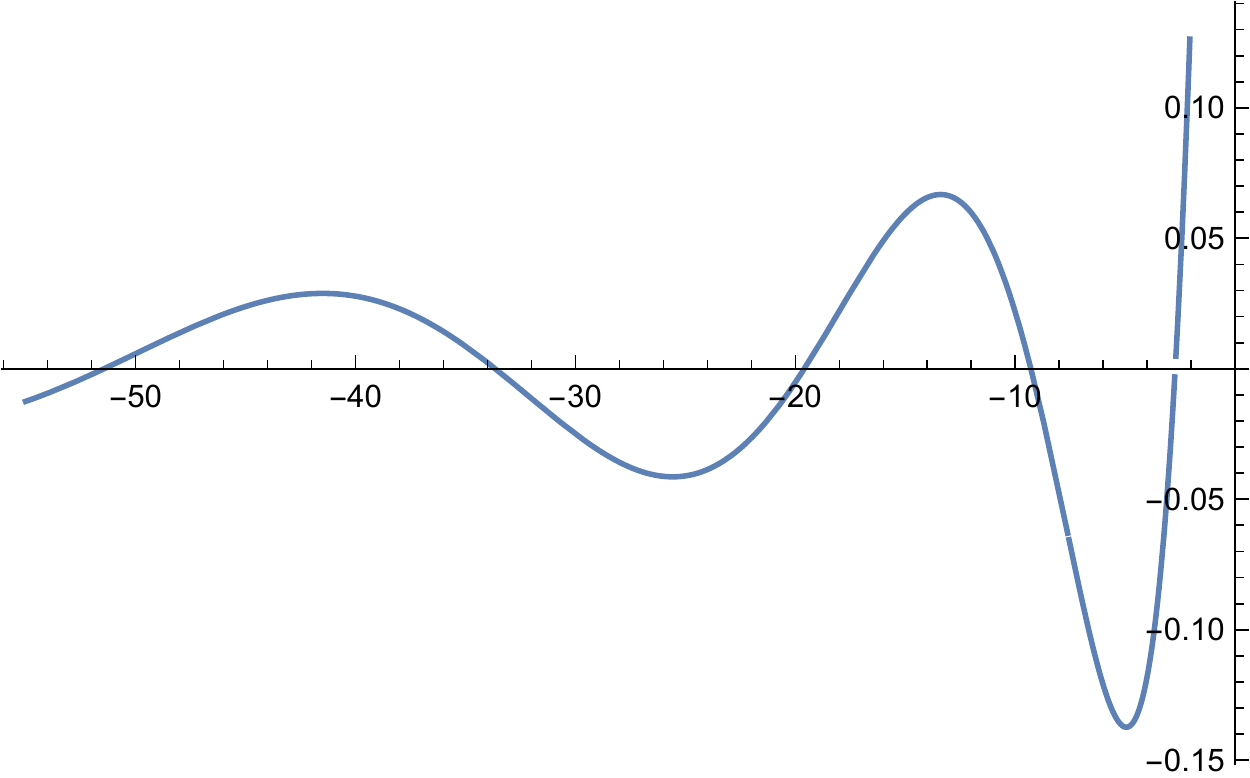}\caption{$\lambda\leq -1$}\label{fig:2a}
\end{subfigure}
\qquad \qquad
\begin{subfigure}[H]{0.4\textwidth}
\includegraphics[width=\textwidth]{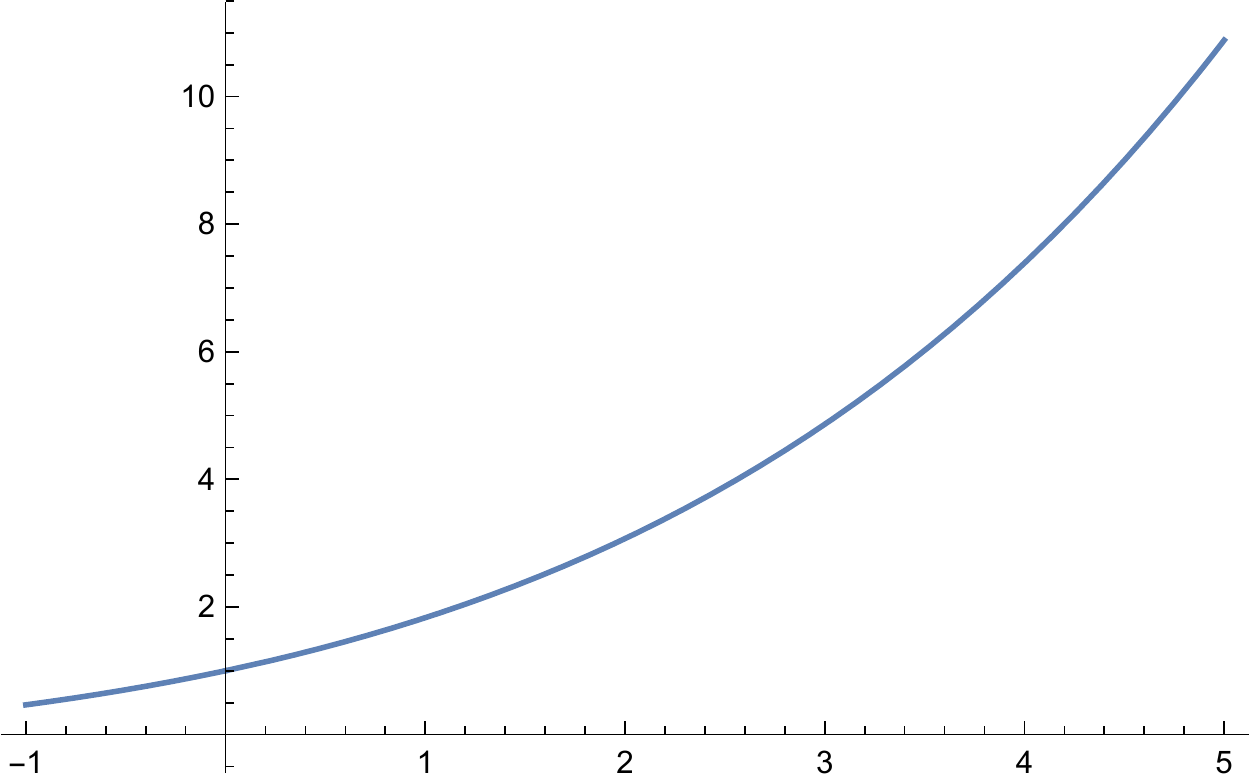}\caption{ $\lambda\geq -1$}\label{fig:2b}
\end{subfigure}
\caption{Graph of $f(\lambda)=e^{-\sqrt{\lambda} }\ {}_1F_1(1+\tfrac{1}{2}\sqrt{\lambda} ~\!;2~\!;2\sqrt{\lambda} )$  as a function of $\lambda$. At $\lambda=0$ we have $f=1$. The graph is again cut in two, at $\lambda=-1$ where there is no discontinuity. Importantly, all the zeros of $f$ are located at negative $\lambda$.
}\label{fig:2}
\end{center}
\end{figure}

For completeness, let us now consider a warp factor corresponding to a $D_6$-brane source located at the origin
\eq{\label{6}
H\simeq H_1+\frac{C}{r} ~,}
with $H_1$, $C$  positive constants. The differential equation \eqref{1} then reduces to
\eq{
\left[r^2 \frac{\d^2}{\d r^2}+2r\frac{\d}{\d r}-\lambda(H_1 r^2+C r)\right]\psi(r)=0
~.}
As before, we may set $H_1$, $C=1$, without loss of generality. Two linearly independent solutions are given by
\eq{
e^{-\sqrt{\lambda} ~\!r}\ {}_1F_1(1+\tfrac{1}{2}\sqrt{\lambda} ~\!;2~\!;2\sqrt{\lambda} ~\!r)~;~~~
e^{-\sqrt{\lambda} ~\!r}\ U(1+\tfrac{1}{2}\sqrt{\lambda} ~\!;2~\!;2\sqrt{\lambda} ~\!r)
~.
}
As before, it can be seen that the first solution is real-valued for any $\lambda\in\mathbb{R}$. To obtain a discrete spectrum for $\lambda$, we again impose the following ``separated'' boundary conditions
\eq{
\psi(0)= \psi(r_0)=0
~,}
for some fixed radial distance $r_0>0$, which we may choose to be $r_0=1$ for simplicity. The first boundary condition implies $\psi(r)\propto {}_1F_1$.   Figure \ref{fig:2} depicts $f(\lambda)=e^{-\sqrt{\lambda} ~\!r_0}\ {}_1F_1(1+\tfrac{1}{2}\sqrt{\lambda} ~\!;2~\!;2\sqrt{\lambda} ~\!r_0)$  as a function of $\lambda$, for $r_0 =1$.
The spectrum of $\lambda$ is then given by the zeros of the function $f$. We note that this time, there are no zeros for positive $\lambda$, i.e.~the spectrum of $\tilde{M}^2$ is strictly positive.

In the $D_p$-brane case, these results are independent of the value of $r_0$. This is related to the fact that the warp factor is positive over the whole domain of definition $[0,r_0]$ of the differential problem, for any value of $r_0$. We conclude on the absence of tachyonic eigenmodes in that case.

\section{Numerical details}\label{ap:num}

\subsection{Making use of the symmetries}

In this appendix, we come back to the numerical method used to determine the spectrum. Let us recall that we initially want to solve \eqref{master}, that is a set of $n$ equations. As mentioned in section \ref{sec:num}, one can use the symmetries of the problem to drastically reduce the size of the system, by merely working with the points $\vm \in \tilde{\Gamma}$ which are non-equivalent under the action of the symmetry group $G$. This is because in \eqref{master}, the norm $|\vn|^2$ is naturally invariant under $G$, and the convolution product is also invariant whenever \eqref{Hsym} and \eqref{psisym} are satisfied (i.e. when we consider the first Kaluza--Klein mode). Indeed, $\forall g \in G$, we have
\beq
\sum_{{\bm m} \in \mathbb{Z}^d} c_{{\bm m}} \, d_{{\bm n} - {\bm m}} = \sum_{{\bm m}'=g^{-1} \cdot {\bm m} \in \mathbb{Z}^d} c_{g\cdot {\bm m'}} \, d_{{\bm n} - g\cdot {\bm m'}} =\sum_{{\bm m'} \in \mathbb{Z}^d} c_{{\bm m'}} \, d_{g^{-1}\cdot{\bm n} -  {\bm m'}} \ ,
\eeq
so acting on ${\bm n}$ in that sum also leaves it invariant. In this case, the system reduces to only $\tilde{n}$ equations
	\begin{equation}
	\frac{|\vn|^2}{\mu_1^2} c_{\vn} - \sum_{\vm \in \Gamma} c_{\vm} \, d_{\vn - \vm} \,  = 0 \,, \qquad \forall \vn \in \tilde{\Gamma} \,.
\end{equation}
We also make use of \eqref{Hsym} and \eqref{psisym} to only consider $c_{\vm}$ and $d_{\vm}$ for $\vm \in \tilde{\Gamma}$, leaving us with only $\tilde{n}$ unknowns $c_{\vm}$ (in addition to $\mu_1$) and only $\tilde{n}$ Fourier coefficients $d_{\vm}$ to compute. To that end, we need to choose a representative $R(\vm) \in \tilde{\Gamma}$ of the equivalence class $[\vm]$ defined by the relation $\vm \sim \vn \Leftrightarrow \vm = g \cdot \vn$ (this amounts to define what we use as $\tilde{\Gamma}$).

Let us note that, in $d$ dimensions, the hyperoctahedral group $G$ can be represented as the signed symmetric group of permutations of a set $\left\{n^1, ..., n^d \right\}$. This provides a way to readily implement its action on points $\vn \in \Gamma$: the orbit of one point is given by the set of possibilities of permuting its components and flipping their signs. Thus, it suffices to sort (say in decreasing order) the absolute value of the components of $\vm$ to pick a representative $R(\vm)$, as hinted around Figure \ref{fig:lattice}. This algorithmic definition of the representative has the advantage to be simple to implement in a computer. For instance in Mathematica, one has
\begin{equation}
	\texttt{R[n\_] := ReverseSort[Abs[n]]} \,.
\end{equation}
With this at hand, one finally recovers \eqref{syst}, that is the set of equations to be solved.

\subsection{Bijective map}

In order to manipulate the elements $\vn$ of $\Gamma$ (or $\tilde{\Gamma}$), it is useful to assign them a label, that is an integer index. This way, the coefficients $d$ and $c$ have an index which is an integer instead of a vector, and this allows to organize the equations and put them in an appropriate matrix form, as was done in \eqref{linear}. We thus need to introduce a one-to-one map,
\begin{equation}
	F \; : \tilde{\Gamma} \longrightarrow \left\{1, 2, ..., \tilde{n}  \right\} \,.
\end{equation}
To implement such a map, one can do the following. Consider a vector $\vn \in \tilde{\Gamma}$ (for concreteness we focus on $d=2$, but everything can be generalized trivially): following the previous subsection, one can take its first entry $n^1$ to be positive, so it goes from $0$ to $\left[r \right]$, where $\left[\cdot \right]$ denotes the integer part. Then for each of these possibilities, $n^2$ can run from $0$  to $\max \left\{n^1, \sqrt{r^2 - (n^1)^2} \right\}$. We then use this construction of $\tilde{\Gamma}$ to label its elements: we start with the origin (indexed 1), then we increase $n^1$, there are now 2 possibilities for $n^2$ (indexed 2 and 3 in order of increasing $n^2$), then we increase  $n^1$ again, and so on.

In order to implement the inverse map $F^{-1}$, we construct a $d$-dimensional array $\left[r \right] \times \left[r \right]\times... \times \left[r \right]$, denoted \texttt{F}. Whenever a point $\vn$ is added to $\tilde{\Gamma}$ via the above algorithm, its label is added at the entry of \texttt{F} that corresponds to the coordinates of $\vn$ (modulo a shift of 1 because indices start at 1 in Mathematica). Let us take an example: in $d=2$, and say $r=3$, \texttt{F} is $3 \times 3$ matrix. The point $(0,0)$ is labeled 1, so the entry $\texttt{F}_{11}$ takes the value 1. The next point $(1,0)$ is labeled 2, so $\texttt{F}_{21}$ takes the value 2. The next point $(1,1)$ is labeled 3, so $\texttt{F}_{22}$ takes the value 3, and so on. The advantage of this approach is that both $F$ and its inverse $F^{-1}$ are executed in constant time $\mathcal{O}(1)$.

\section{Tachyonic spectrum}\label{ap:old}

In this appendix, we provide for completeness the Kaluza--Klein spectrum obtained for $d=1,2,3$ using the same method as in \cite{Andriot:2019hay}, namely the determinant method explained around \eqref{det} and a constant $H_0$ given by \eqref{constantH0}. The spectrum is improved with respect to \cite{Andriot:2019hay} in terms of precision and number of modes thanks to the bijective map described at the beginning of section \ref{sec:num}. We give in tables \ref{tab:tachd1}-\ref{tab:tachd3} the spectrum of the first modes of the tower, together with the sample specifications (see section \ref{sec:num}).

A crucial difference with the results of section \ref{sec:spectrum} is the presence here of tachyons for $d\geq 2$,\footnote{Given that there is no divergence at the source for $d=1$, one can find a constant $H_0$ that makes $H\geq 0$ everywhere, as done in \cite{Andriot:2019hay}, thus avoiding any tachyon in that case.} as indicated in tables \ref{tab:tachd2} and \ref{tab:tachd3}. The resolution of the eigenmode equation is indeed performed here on the complete torus $\mathbb{T}^d$, including the negative region where $H<0$. As discussed in section \ref{sec:neg} and appendix \ref{ap:noncomp}, this region is responsible for tachyonic modes in the spectrum. The solution proposed in section \ref{sec:D} is then to solve the equation on a restricted domain ${\cal D}$ that excludes the negative region. One can compare the resulting spectrum in section \ref{sec:spectrum} to the one here, at the same $L/l_s$, and note the absence of tachyon in section \ref{sec:spectrum}.

Let us recall that we only consider a finite set of eigenmodes, because of the truncation made to solve the problem, as discussed in section \ref{sec:num}. When varying $L/l_s$, we can follow the evolution of the mass of each mode (or multiplet of modes with certain degeneracies). We note that the tachyons only become apparent to us at low $L/l_s$. In addition, they correspond to modes which were among most massive at higher $L/l_s$, and which disappear when lowering $L/l_s$. This is made manifest in tables \ref{tab:tachd2} and \ref{tab:tachd3}. As already mentioned in section \ref{sec:neg}, we understand this phenomenon as follows: the eigenmodes need to have a small enough wavelength, i.e.~a high enough mass, to probe the negative region. When $L/l_s$ becomes small, the region becomes large and modes at high mass within our truncation then have the appropriate wavelength to become tachyonic, through the integral \eqref{intHtachyon}. In other words, tachyons may always be present in the spectrum, but for high $L/l_s$, they would lie outside our truncation. It would be interesting to test further this interpretation.\footnote{An estimate of the wavelength is given by $1/N=1/|{\bm m}|$. The criterion for a tachyon to appear would then be that the wavelength is shorter or of the same order as the typical size of the negative region, i.e.~$1/N \lesssim 2 l_s/L$. We verify approximately in tables \ref{tab:tachd2} and \ref{tab:tachd3} this inequality $L/(2l_s) \lesssim N$ for the tachyonic mode. In addition, we also note that $N \approx r$, since the tachyon is among the highest modes of the truncation. Given a value of $r$, we deduce that a tachyon will appear for $L/l_s \lesssim 2 r$.}

\begin{table}[H]
	\begin{center}
		\begin{tabular}{|c||c|c||c|c|}
			\hline
			$N$ & \multicolumn{2}{c||}{$1$} & \multicolumn{2}{c|}{$2$} \\
			\hline
			$a/s$  & $a$ & $s$ & $a$ & $s$ \\
			\hline
			\hline
			$\mu_N$& 0.9800 & 1.1400 & 2.037 & 2.205 \\
			\hline
			$f_N$&  0.9800 & 1.1400 & 1.018 & 1.102 \\
			\hline
		\end{tabular}
		\caption{Spectrum of the first Kaluza--Klein modes for $d=1$, with eigenvalue $\mu_N$ and deviation $f_N$ from the standard spectrum, using the (improved) method of \cite{Andriot:2019hay}. Sample specifications: $r=20,\, n = 41$.}\label{tab:tachd1}
	\end{center}
\end{table}

\begin{table}[H]
	\begin{center}
		\begin{tabular}{|c||c||c|c|c||c|c|c|}
			\hline
			& $N$ & \multicolumn{3}{c||}{$1$} & \multicolumn{3}{c|}{$\sqrt{2}$} \\
			\cline{2-8}
			$L/l_s$ & $a/s$ &$s(1)$& $a(2)$ & $s(1)$&$s(1)$ & $a(2)$ & $s(1)$ \\
			\hline
			\hline
			& $\mu_N$ & 0.7537 & 0.7944 & 0.9709 & 1.154 & 1.245 & 1.340  \\
			\cline{2-8}
			$9$ & $f_N$   & 0.8914 & 0.9395 & 1.148 & 0.9651 & 1.041 & 1.121 \\
			\hline
			\hline
			& $\mu_N$ & 0.7378 & 0.7798 & 0.9435 & 1.131 & 1.213 & 1.305  \\
			\cline{2-8}
			$10$ & $f_N$   & 0.8932 & 0.9441 &  1.142 & 0.9681 & 1.039 & 1.117 \\
			\hline
			\hline
			& $\mu_N$ & 0.5439 & 0.5749 & 0.6260 & 0.8192 & 0.8350 & 0.8692  \\
			\cline{2-8}
			$10^2$ & $f_N$   & 0.9312 & 0.9843 &  1.072 & 0.9918 & 1.011 & 1.052 \\
			\hline
			\hline
			& $\mu_N$ & 0.4539 & 0.4735 & 0.4995 & 0.6719 & 0.6776 & 0.6925  \\
			\cline{2-8}
			$10^3$ & $f_N$   & 0.9519 & 0.9929 &  1.048 & 0.9964 & 1.005 & 1.027 \\
			\hline
		\end{tabular}
		\quad
		\begin{tabular}{|c|}
			\hline
			$2\sqrt{5}$\\
            \hline
			Tachyon\\
			\hline
			\hline
			56.51 $\i$ \\
			\hline
			/
			\\
			\hline
			\hline
			10.54 \\
			\hline
			2.852
			\\
			\hline
			\hline
			2.920 \\
			\hline
			1.118
			\\
			\hline
			\hline
			2.247 \\
			\hline
			1.054
			\\
			\hline
		\end{tabular}
		\caption{Spectrum of the first Kaluza--Klein modes for $d=2$, with eigenvalue $\mu_N$ and deviation $f_N$ from the standard spectrum, according to the value of $L/l_s$, using the (improved) method of \cite{Andriot:2019hay}. Sample specifications: $r=4.5,\, n = 69$.}\label{tab:tachd2}
	\end{center}
\end{table}

\begin{table}[H]
	\begin{center}
		\hspace{-2cm}
		\begin{tabular}{|c||c||c|c|c||c|c|c|c|c|}
			\hline
			& $N$ & \multicolumn{3}{c||}{$1$} & \multicolumn{5}{c|}{$\sqrt{2}$} \\
			\cline{2-10}
			$L/l_s$ & $a/s$ &$s(1)$& $a(3)$ & $s(2)$&$s(1)$ & $a(3)$ & $s(3)$  & $a(3)$& $s(2)$ \\
			\hline
			\hline
			& $\mu_N$ & 0.6681 & 0.7627 & 1.049 & 1.350 & 1.153 & 1.237 & 1.464 & 1.526  \\
			\cline{2-10}
			$7$ & $f_N$   & 0.7052 & 0.805 & 1.108 & 1.007 & 0.8607 & 0.9236 & 1.093 & 1.139 \\
			\hline
			\hline
			& $\mu_N$ & 0.5899 & 0.6969 & 0.8729 & 1.118 & 1.027 & 1.028 & 1.205 & 1.275  \\
			\cline{2-10}
			$10$ & $f_N$   & 0.7442 & 0.8792 & 1.101 & 0.9971 & 0.916 & 0.9175 & 1.075 & 1.138 \\
			\hline
			\hline
			& $\mu_N$ & 0.2433 & 0.2504 & 0.2538 & 0.3444 & 0.3507 & 0.3542 & 0.3575 & 0.3607  \\
			\cline{2-10}
			$10^2$ & $f_N$   & 0.9704 & 0.9988 &  1.013 & 0.9714 & 0.9892 & 0.9991 & 1.008 & 1.017 \\
			\hline
			\hline
			& $\mu_N$ & 0.07906 & 0.07927 & 0.07937 & 0.1117 & 0.1120 & 0.1121 & 0.1122 & 0.1123 \\
			\cline{2-10}
			$10^3$ & $f_N$   & 0.9974 & 1.000 & 1.001 & 0.9966 & 0.9991 & 1.000 & 1.001 & 1.002 \\
			\hline
		\end{tabular}
		\ \
		\begin{tabular}{|c|}
			\hline
			$\sqrt{6}$\\
            \hline
			Tachyon\\
			\hline
			\hline
			5.237 $\i$ \\
			\hline
			/
			\\
			\hline
			\hline
			3.019
			\\
			\hline
			1.555
			\\
			\hline
			\hline
			0.6231
			\\
			\hline
			1.015
			\\
			\hline
			\hline
			0.1944
			\\
			\hline
			1.001
			\\
			\hline
		\end{tabular}
		\caption{Spectrum of the first Kaluza--Klein modes for $d=3$, with eigenvalue $\mu_N$ and deviation $f_N$ from the standard spectrum, according to the value of $L/l_s$, using the (improved) method of \cite{Andriot:2019hay}. Sample specifications: $r=2.6,\, n = 81$.}\label{tab:tachd3}
	\end{center}
\end{table}

We finally add a word on the degeneracies of each level. In the limit of large $L/l_s$, one recovers the standard Kaluza--Klein spectrum \eqref{st} on the torus $\mathbb{T}^d$. There, the eigenmodes are the Fourier modes, and the label $N$ of each level is given by the norm of the vector ${\bm m} \in \mathbb{Z}^d$ of the eigenmode. The spectrum is thus discrete and corresponds to points in $\mathbb{Z}^d$ obtained when increasing the norm. Because of the symmetries of the lattice, the spectrum is degenerate; the degeneracy corresponds to the number of lattice points having the same norm. An illustration is provided in Figure \ref{fig:lattice}. Knowing the exact degeneracy and the mass gap is actually a non-trivial question. It corresponds to a ``generalized Gauss circle problem'' in arbitrary dimension $d$. For $d=2$, it is the problem of determining how many integer lattice points there are in a circle centered at the origin and with radius $r$ as in Figure \ref{fig:lattice}. In that case, the first levels are given by $N=1, \sqrt{2}, 2, \sqrt{5}, ...$, while the degeneracies are respectively given by ${\cal D}_N=4, 4, 4, 8, ...$ . This knowledge is interesting to us, as it allows to classify the masses for the non-standard spectrum, for which the degeneracy is (partially) lifted: see Figure \ref{fig:spectred3}. Indeed, because of the different mass values, it is then unclear which one corresponds to a given level $N$. For $d=2$, we now know that the first $4$ masses correspond to $N=1$, then the next $4$ masses correspond to $N=\sqrt{2}$, etc.

It is easy to determine the degeneracy of levels due to vectors ${\bm m}=(\pm 1,...,\pm 1,0,...,0)$, i.e.~$k$ unit vectors (or their opposite) in $d$ dimensions: those give $N=\sqrt{k}$. In that case, provided that no vector of a different kind contributes to that level, one gets the degeneracy
\begin{equation}
	{\cal D}_{\sqrt{k}} = {d\choose k} \times 2^k \,, \qquad 1\leq k \leq d \,.
\end{equation}
This provides the first $k$ levels for $k\leq d \leq 3$, and the first 3 levels for $d\geq 4$. We thus give the degeneracies of the first 3 levels in table \ref{Tabledegen}. They are useful to identify the eigenmodes in the spectra: the one given above or that of section \ref{sec:spectrum}.
\begin{table}[h]
	\begin{center}
		\begin{tabular}{|c|c|c|c|c|c|}
			\cline{4-6}
			\multicolumn{3}{c|}{} & \multicolumn{3}{c|}{Level} \\
			\cline{4-6}
			\multicolumn{3}{c|}{}  & 1 & 2 & 3 \\
			\hline
			\multirow{12}{*}{\begin{sideways} Dimension $d$~ \end{sideways}} &  & $N$ & 1 & 2 & 3 \\
			\cline{3-6}
			& 1 & ${\cal D}_N$ & 2 & 2 & 2  \\
			\cline{2-6}
			&  & $N$ & 1 & $\sqrt{2}$ & 2 \\
			\cline{3-6}
			& 2 & ${\cal D}_N$ & 4 & 4 & 4  \\
			\cline{2-6}
			&  & $N$ & 1 & $\sqrt{2}$ & $\sqrt{3}$ \\
			\cline{3-6}
			& 3 & ${\cal D}_N$ & 6 & 12 & 8 \\
			\cline{2-6}
			&  & $N$ & 1 & $\sqrt{2}$ & $\sqrt{3}$ \\
			\cline{3-6}
			& 4 & ${\cal D}_N$ & 8  & 24 & 32  \\
			\cline{2-6}
			&  & $N$ & 1 & $\sqrt{2}$ & $\sqrt{3}$ \\
			\cline{3-6}
			& 5 & ${\cal D}_N$ & 10 & 40 & 80  \\
			\cline{2-6}
			&  & $N$ & 1 & $\sqrt{2}$ & $\sqrt{3}$ \\
			\cline{3-6}
			& 6 & ${\cal D}_N$ & 12 & 60 & 160  \\
			\hline
		\end{tabular}
	\end{center}
	\caption{Label $N$ and degeneracy ${\cal D}_N$ of the first 3 Kaluza--Klein levels for each dimension $d$.}
	\label{Tabledegen}
\end{table}

\end{appendix}

\newpage

\providecommand{\href}[2]{#2}\begingroup\raggedright
\endgroup

\end{document}

%% file: warpfactor.pdf_tex
\begingroup%
  \makeatletter%
  \providecommand\color[2][]{%
    \errmessage{(Inkscape) Color is used for the text in Inkscape, but the package 'color.sty' is not loaded}%
    \renewcommand\color[2][]{}%
  }%
  \providecommand\transparent[1]{%
    \errmessage{(Inkscape) Transparency is used (non-zero) for the text in Inkscape, but the package 'transparent.sty' is not loaded}%
    \renewcommand\transparent[1]{}%
  }%
  \providecommand\rotatebox[2]{#2}%
  \newcommand*\fsize{\dimexpr\f@size pt\relax}%
  \newcommand*\lineheight[1]{\fontsize{\fsize}{#1\fsize}\selectfont}%
  \ifx\svgwidth\undefined%
    \setlength{\unitlength}{402.72531104bp}%
    \ifx\svgscale\undefined%
      \relax%
    \else%
      \setlength{\unitlength}{\unitlength * \real{\svgscale}}%
    \fi%
  \else%
    \setlength{\unitlength}{\svgwidth}%
  \fi%
  \global\let\svgwidth\undefined%
  \global\let\svgscale\undefined%
  \makeatother%
  \begin{picture}(1,0.57244842)%
    \lineheight{1}%
    \setlength\tabcolsep{0pt}%
    \put(0,0){\includegraphics[width=\unitlength,page=1]{warpfactor.pdf}}%
    \put(0.4774398,0.59514147){\color[rgb]{0,0,0}\makebox(0,0)[lt]{\lineheight{1.25}\smash{\begin{tabular}[t]{l}\fontsize{20pt}{1em}$H$\end{tabular}}}}%
    \put(1.02140306,0.21461366){\color[rgb]{0,0,0}\makebox(0,0)[lt]{\lineheight{1.25}\smash{\begin{tabular}[t]{l}\fontsize{20pt}{1em}$\sigma$\end{tabular}}}}%
    \put(0.93187347,0.17021677){\color[rgb]{0,0,0}\makebox(0,0)[lt]{\lineheight{1.25}\smash{\begin{tabular}[t]{l}\fontsize{20pt}{1em}$\frac{1}{2}$\end{tabular}}}}%
    \put(0.0007567,0.17021677){\color[rgb]{0,0,0}\makebox(0,0)[lt]{\lineheight{1.25}\smash{\begin{tabular}[t]{l}\fontsize{20pt}{1em}$-\frac{1}{2}$\end{tabular}}}}%
    \put(0.65199521,0.15102908){\color[rgb]{0,0,0}\makebox(0,0)[lt]{\lineheight{1.25}\smash{\begin{tabular}[t]{l}\fontsize{20pt}{1em}$\lambda$\end{tabular}}}}%
    \put(0,0){\includegraphics[width=\unitlength,page=2]{warpfactor.pdf}}%
    \put(0.85224822,0.27786008){\color[rgb]{0,0,0}\makebox(0,0)[lt]{\lineheight{1.25}\smash{\begin{tabular}[t]{l}\fontsize{15pt}{1em}$l_s/L$\end{tabular}}}}%
  \end{picture}%
\endgroup%

%% file: domain.pdf_tex
\begingroup%
  \makeatletter%
  \providecommand\color[2][]{%
    \errmessage{(Inkscape) Color is used for the text in Inkscape, but the package 'color.sty' is not loaded}%
    \renewcommand\color[2][]{}%
  }%
  \providecommand\transparent[1]{%
    \errmessage{(Inkscape) Transparency is used (non-zero) for the text in Inkscape, but the package 'transparent.sty' is not loaded}%
    \renewcommand\transparent[1]{}%
  }%
  \providecommand\rotatebox[2]{#2}%
  \newcommand*\fsize{\dimexpr\f@size pt\relax}%
  \newcommand*\lineheight[1]{\fontsize{\fsize}{#1\fsize}\selectfont}%
  \ifx\svgwidth\undefined%
    \setlength{\unitlength}{486.18264987bp}%
    \ifx\svgscale\undefined%
      \relax%
    \else%
      \setlength{\unitlength}{\unitlength * \real{\svgscale}}%
    \fi%
  \else%
    \setlength{\unitlength}{\svgwidth}%
  \fi%
  \global\let\svgwidth\undefined%
  \global\let\svgscale\undefined%
  \makeatother%
  \begin{picture}(1,0.99094701)%
    \lineheight{1}%
    \setlength\tabcolsep{0pt}%
    \put(0,0){\includegraphics[width=\unitlength,page=1]{domain.pdf}}%
    \put(0.04139164,0.85093429){\color[rgb]{0,0,0}\makebox(0,0)[lt]{\lineheight{1.25}\smash{\begin{tabular}[t]{l}\fontsize{20pt}{1em}$\frac{1}{2}$\end{tabular}}}}%
    \put(0.85722544,0.0319996){\color[rgb]{0,0,0}\makebox(0,0)[lt]{\lineheight{1.25}\smash{\begin{tabular}[t]{l}\fontsize{20pt}{1em}$\frac{1}{2}$\end{tabular}}}}%
    \put(0.07628947,1.00642048){\color[rgb]{0,0,0}\makebox(0,0)[lt]{\lineheight{1.25}\smash{\begin{tabular}[t]{l}\fontsize{20pt}{1em}$\sigma^2$\end{tabular}}}}%
    \put(1.01234542,0.08248512){\color[rgb]{0,0,0}\makebox(0,0)[lt]{\lineheight{1.25}\smash{\begin{tabular}[t]{l}\fontsize{20pt}{1em}$\sigma^1$\end{tabular}}}}%
    \put(0.42735835,0.44143689){\color[rgb]{0,0,0}\makebox(0,0)[lt]{\lineheight{1.25}\smash{\begin{tabular}[t]{l}\fontsize{40pt}{1em}$\mathcal{D}$\end{tabular}}}}%
    \put(0,0){\includegraphics[width=\unitlength,page=2]{domain.pdf}}%
    \put(0.25606685,0.80688859){\color[rgb]{0,0,0}\makebox(0,0)[lt]{\lineheight{1.25}\smash{\begin{tabular}[t]{l}\fontsize{20pt}{1em}$l_s/L$\end{tabular}}}}%
    \put(0.77953405,0.2613376){\color[rgb]{0,0,0}\makebox(0,0)[lt]{\lineheight{1.25}\smash{\begin{tabular}[t]{l}\fontsize{20pt}{1em}$l_s/L$\end{tabular}}}}%
    \put(0.11479632,0.049504){\color[rgb]{0,0,0}\makebox(0,0)[lt]{\lineheight{1.25}\smash{\begin{tabular}[t]{l}\fontsize{20pt}{1em}$0$\end{tabular}}}}%
  \end{picture}%
\endgroup%

%% file: Lattice.pdf_tex
\begingroup%
  \makeatletter%
  \providecommand\color[2][]{%
    \errmessage{(Inkscape) Color is used for the text in Inkscape, but the package 'color.sty' is not loaded}%
    \renewcommand\color[2][]{}%
  }%
  \providecommand\transparent[1]{%
    \errmessage{(Inkscape) Transparency is used (non-zero) for the text in Inkscape, but the package 'transparent.sty' is not loaded}%
    \renewcommand\transparent[1]{}%
  }%
  \providecommand\rotatebox[2]{#2}%
  \newcommand*\fsize{\dimexpr\f@size pt\relax}%
  \newcommand*\lineheight[1]{\fontsize{\fsize}{#1\fsize}\selectfont}%
  \ifx\svgwidth\undefined%
    \setlength{\unitlength}{382.82794622bp}%
    \ifx\svgscale\undefined%
      \relax%
    \else%
      \setlength{\unitlength}{\unitlength * \real{\svgscale}}%
    \fi%
  \else%
    \setlength{\unitlength}{\svgwidth}%
  \fi%
  \global\let\svgwidth\undefined%
  \global\let\svgscale\undefined%
  \makeatother%
  \begin{picture}(1,0.98276877)%
    \lineheight{1}%
    \setlength\tabcolsep{0pt}%
    \put(0,0){\includegraphics[width=\unitlength,page=1]{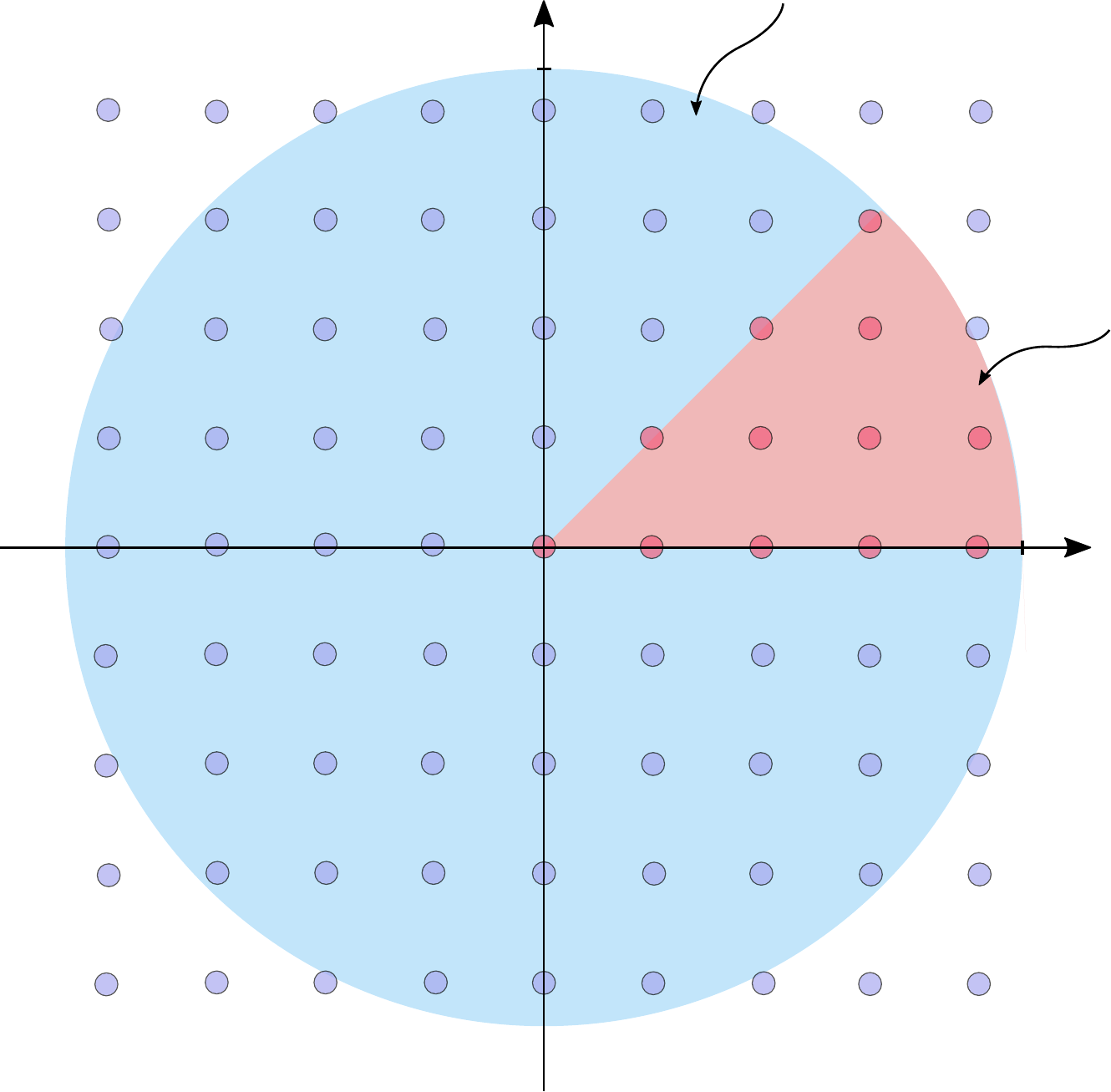}}%
    \put(0.69772106,0.99956109){\color[rgb]{0,0,0}\makebox(0,0)[lt]{\lineheight{1.25}\smash{\begin{tabular}[t]{l}\fontsize{30pt}{1em}$\Gamma$\end{tabular}}}}%
    \put(1.01379774,0.69107773){\color[rgb]{0,0,0}\makebox(0,0)[lt]{\lineheight{1.25}\smash{\begin{tabular}[t]{l}\fontsize{30pt}{1em}$\tilde{\Gamma}$\end{tabular}}}}%
    \put(1.00407189,0.48014693){\color[rgb]{0,0,0}\makebox(0,0)[lt]{\lineheight{1.25}\smash{\begin{tabular}[t]{l}\fontsize{20pt}{1em}$n^1$\end{tabular}}}}%
    \put(0.47245599,1.00378761){\color[rgb]{0,0,0}\makebox(0,0)[lt]{\lineheight{1.25}\smash{\begin{tabular}[t]{l}\fontsize{20pt}{1em}$n^2$\end{tabular}}}}%
    \put(0.90834262,0.51653035){\color[rgb]{0,0,0}\makebox(0,0)[lt]{\lineheight{1.25}\smash{\begin{tabular}[t]{l}\fontsize{20pt}{1em}$r$\end{tabular}}}}%
    \put(0.51736672,0.91003053){\color[rgb]{0,0,0}\makebox(0,0)[lt]{\lineheight{1.25}\smash{\begin{tabular}[t]{l}\fontsize{20pt}{1em}$r$\end{tabular}}}}%
  \end{picture}%
\endgroup%